\RequirePackage{fix-cm} % Fix LaTeX2e bugs.
\documentclass[a4paper, twoside, reqno, 12pt, dvips]{amsart}
\usepackage{fixltx2e}     % Fix LaTeX2e bugs.

\usepackage{amssymb}
\usepackage{amsmath}
\usepackage{amsfonts}
\usepackage{verbatim}
\usepackage{mathrsfs, dsfont}

\usepackage{a4wide}
\usepackage{xspace}

\usepackage{indentfirst}
\usepackage{graphicx}
\usepackage{subfigure}
\usepackage[section]{placeins}
\usepackage[bf, up, hang]{caption}

\usepackage{epsfig}
\usepackage{color}
\definecolor{oneblue}{rgb}{0.0, 0.0, 0.85}
\definecolor{darkgrey}{rgb}{0.273, 0.281, 0.30}

\usepackage[colorlinks,
            urlcolor=oneblue,
            linkcolor=oneblue,
            citecolor=oneblue,
            bookmarksopen=false,
            pagebackref]{hyperref}

\usepackage[compact]{titlesec}

\titleformat{\section}{\normalfont\Large\bfseries\sffamily\center\color{darkgrey}}{\thesection.}{0.5em}{}{}
\titleformat{\subsection}{\normalfont\large\bfseries\sffamily\color{darkgrey}}{\thesubsection.}{0.4em}{}{}
\titleformat{\subsubsection}{\normalfont\normalsize\bfseries\sffamily\color{darkgrey}}{\thesubsubsection.}{0.3em}{}{}

\titlespacing*{\section}{1.0em}{1.0em}{0.8em}[0em]
\titlespacing*{\subsection}{1.0em}{1.0em}{0.8em}[0em]
\titlespacing*{\subsubsection}{1.0em}{0.7em}{0.6em}[0em]

\usepackage{titletoc}

\contentsmargin{0.0em}
\dottedcontents{section}[2.5em]{\addvspace{0.45em}\bfseries}{1.0em}{0pc}
\dottedcontents{subsection}[4.5em]{}{2.6em}{0pc}
\dottedcontents{subsubsection}[5.5em]{}{3.0em}{0pc}

\usepackage{fancyhdr}
\usepackage{lastpage}

\newcommand*\Title{Boussinesq modeling of underwater landslides}
\newcommand*\Authors{D.~Dutykh \& H.~Kalisch}

\usepackage{acronym}
\numberwithin{equation}{section}

\newtheorem{remark}{Remark}

\newcommand{\um}{\bar{\/u}}
\newcommand{\hm}{\bar{\/H}}

\newcommand{\q}{\mathbf{q}}
\newcommand{\V}{\mathbf{V}}
\newcommand{\vm}{\V}
\newcommand{\Q}{\mathbf{Q}}
\newcommand{\W}{\mathbf{W}}
\newcommand{\F}{\mathbb{F}}
\newcommand{\R}{\mathbb{R}}
\newcommand{\A}{\mathbb{A}}
\newcommand{\Z}{\mathbb{Z}}
\newcommand{\U}{\mathbb{U}}
\newcommand{\M}{\mathbb{M}}

\newcommand{\N}{\mathcal{N}}
\newcommand{\Sl}{\mathbf{S}}
\newcommand{\ud}{\mathrm{d}}
\newcommand{\ue}{\mathrm{e}}
\newcommand{\ui}{\mathrm{i}}
\newcommand{\C}{\mathcal{C}}
\newcommand{\Fr}{\mathrm{Fr}}
\newcommand{\It}{\mathcal{I}}
\renewcommand{\S}{\mathbb{S}}

\renewcommand{\O}{\mathcal{O}}
\newcommand{\scal}{\boldsymbol{\cdot}}
\newcommand{\frth}{{\textstyle{3\over4}}}
\newcommand{\half}{{\textstyle{1\over2}}}
\newcommand{\sign}{\mathop{\mathrm{sign}}}
\newcommand{\diag}{\mathop{\mathrm{diag}}}
\newcommand{\minmod}{\mathop{\mathrm{minmod}}}
\newcommand{\pd}[2]{\frac{\partial\, #1}{\partial\/ #2}}
\newcommand{\od}[2]{\frac{\mathrm{d}#1}{\mathrm{d}\/#2}}
\newcommand{\odd}[2]{\frac{\mathrm{d}^2#1}{\mathrm{d}\/#2^2}}
\newcommand{\oddd}[2]{\frac{\mathrm{d}^3#1}{\mathrm{d}\/#2^3}}

\newcommand{\sech}{\mathrm{sech} \,}
\newcommand{\LL}{\mathcal{L}}

\begin{document}

\title[\Title]%
{Boussinesq modeling of surface waves due to underwater landslides}

\author[D.~Dutykh]{Denys Dutykh$^*$}
\address{University College Dublin, School of Mathematical Sciences, Belfield, Dublin 4, Ireland \and LAMA, UMR 5127 CNRS, Universit\'e de Savoie, Campus Scientifique, 73376 Le Bourget-du-Lac Cedex, France}
\email{Denys.Dutykh@ucd.ie}
\urladdr{http://www.denys-dutykh.com/}
\thanks{$^*$ Corresponding author}

\author[H.~Kalisch]{Henrik Kalisch}
\address{Department of Mathematics, University of Bergen, Postbox 7800, 5020 Bergen, Norway}
\email{Henrik.Kalisch@math.uib.no}
\urladdr{http://folk.uib.no/hka002/}

\begin{abstract}
Consideration is given to the influence of an underwater landslide on waves at the surface of a shallow body of fluid. The equations of motion which govern the evolution of the barycenter of the landslide mass include various dissipative effects due to bottom friction, internal energy dissipation, and viscous drag. The surface waves are studied in the Boussinesq scaling, with time-dependent bathymetry. A numerical model for the Boussinesq equations is introduced which is able to handle time-dependent bottom topography, and the equations of motion for the landslide and surface waves are solved simultaneously.

The numerical solver for the Boussinesq equations can also be restricted to implement a shallow-water solver, and the shallow-water and Boussinesq configurations are compared. A particular bathymetry is chosen to illustrate the general method, and it is found that the Boussinesq system predicts larger wave run-up than the shallow-water theory in the example treated in this paper. It is also found that the finite fluid domain has a significant impact on the behavior of the wave run-up.

\bigskip
\noindent \textbf{\keywordsname:} Surface waves, Boussinesq model, submarine landslides, wave run-up, tsunami.

\end{abstract}

\maketitle
\tableofcontents
\thispagestyle{empty}

% **********************************************************************************
\section{Introduction}\label{introduction}
% **********************************************************************************

Surface waves originating from sudden perturbations of the bottom topography are often termed tsunamis. Two distinct generation mechanisms of a tsunami are underwater earthquakes, and submarine mass failures. Among the broad class of submarine mass failures, landslides can be characterised as translational failures which travel considerable distances along the bottom profile \cite{Grilli2005, Prior1979}. In the past, the role of landslides and rock falls in the excitation of tsunamis may have been underestimated, as most known occurrences of tsunamis were accredited to seismic activity. However, it is now more accepted that submarine mass failures also contribute to a large portion of tsunamis \cite{Tinti2001}, and recent years have seen a multitude of works devoted to the study of such underwater landslides and the resulting effect on surface waves \cite{Bardet2003, Chubarov2011, Didenkulova2010, Fernandez-Nieto2007, Grilli1999, Grilli2005, Okal2003NEW, OkalSynolakis2003, PoncetCanada2010, Tinti2001}. As suggested in \cite{Fritz2007}, it is possible for underwater landslides and earthquakes to act in tandem, and produce very large surface waves 

A natural question to ask is whether the effect of underwater landslides on surface waves can be such that they may pose a danger for civil engineering structures located near the shore. Consequently, one important issue is the wave action and in particular the run-up and draw-down at beaches in the vicinity of the landslide. While the draw-down itself may not pose a threat, one consequence of a large draw-down can be the amplification of the run-up of the following positive wave crest \cite{Dutykh2011, Synolakis1996}.

There have been many numerical and a few experimental studies devoted to this subject, but it is generally difficult to include many of the complex parameters and dependencies of a realistic landslide into a physical model. Therefore, most workers attempt to distill the problem to a model setup where many effects such as turbulence and sedimentation are disregarded. For example, Grilli and Watts \cite{Grilli2005} study tsunami sensitivity to several landslide parameters in the case of a landslide in a coastal area of an open ocean. In particular, dependence on the landslide shape and the initial depth of the landslide location are studied, and it is found that the landslide with the smallest length produced the largest waveheight and run-up, and that the wave run-up at an adjacent beach is inversely proportional to the initial depth. The work in \cite{Grilli2005} relies on integrating the full water-wave equations using an irrotational boundary-element code, and using an open boundary with transmission conditions \cite{Grilli2001,GrilliDias2010}. While most works have considered a given dynamics for the landslide, the bottom motion in \cite{Grilli2005} is described by an ordinary differential equation similar to the one used here. Thus the motion of the landslide is computed using a differential equation derived from first principles using Newtonian mechanics. However to expedite comparison with experiments, the landslide in \cite{Grilli2005} is considered moving on a straight inclined bottom with constant slope.

More recently, Khakimzyanov and Shokina \cite{KhakimzyanovG.S.Shokina2010}, and Chubarov \emph{et. al}. \cite{Chubarov2011} have also used a differential equation to find the bottom motion. One major novelty in their work is that the landslide motion is computed on a bottom with an arbitrary shape. The time-dependent bathymetry is then used to drive a numerical solver of the shallow-water equations. An advantage of this approach when compared to \cite{Grilli2005} is the reduced computation time. On the other hand, the description of the wave motion in the shallow-water theory is only approximate, and in particular, one important effect of surface waves, namely the influence of frequency dispersion is neglected. 

\begin{figure}
\centering
\includegraphics[width=0.69\textwidth]{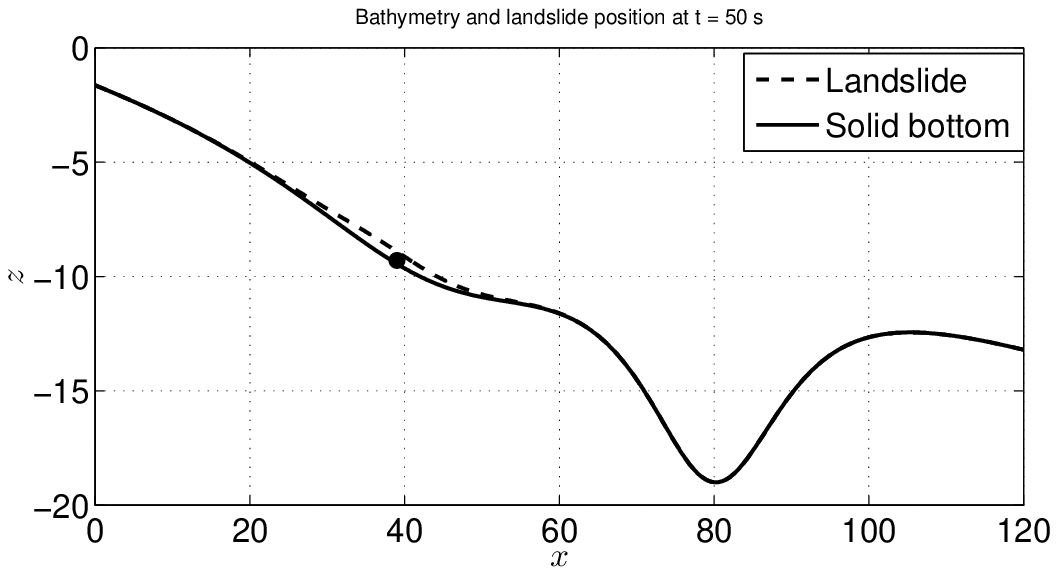}
\caption{\small\em The fixed bathymetry $z = h_0(x)$ and the position of the landslide after $50$ {\sf s}. 
The position of the barycenter is indicated by a black dot.}
\label{SlideBathy}
\end{figure}

The main aim of the current work is to study the dispersive wave generation in a closed basin \cite{Beisel2012} using a more realistic landslide model \cite{Chubarov2005} while keeping the simplicity of the shallow-water approach. To this end, we use the so-called Peregrine system which is a particular case of a general class of model systems which arise in the Boussinesq scaling \cite{bouss}. A common feature of all Boussinesq-type systems is that they allow a simplified study of surface waves in which both nonlinear and dispersive effects are taken into account. In the present case, we need to use a Boussinesq system which can handle complex and time-dependent bottom topography. Such a system was derived by Wu \cite{Wu1987}, and can be used in connection with the dynamic bathymetry. An example of the type of situation considered here is shown in Figure \ref{SlideBathy}, which shows how the bathymetry is given by the combination of a solid bottom, and a landslide profile sliding along the fixed bottom.

We conduct two main experiments. First, a comparison with the shallow-water theory is carried out. Second, the dependence of the tsunami characteristics on the initial depth of the landslide is investigated. The main findings of the present work are that the predictions of the shallow-water and Boussinesq theory are divergent for the cases treated in this paper, and that the effect of a finite fluid domain, such as a river, lake or fjord \cite{PoncetCanada2010} can lead to significantly different behavior when compared to tsunamis on an open ocean (see also \cite{Beisel2012}).
The Boussinesq model in this paper is based on the assumption of an inviscid fluid, and irrotational flow. These are standard assumptions in the study of surface waves, and generally give good results, unless there are strong background currents in the fluid. Another effect which is not taken account of here is the wave resistance on the landslide due to waves created by the motion of the landslide. However, as observed in \cite{Harbitz2006}, this effect is negligible for most realistic cases of underwater landslides. Viscosity {\it is} included in the dynamic model for the landslide as will be shown in the next section. In order to capture the effect of slide deformation during the evolution, a damping term in the equation of motion is included to model the internal friction in the landslide mass. 

The paper is organised in the following way. In Section~\ref{LandslideModel}, the equation of motion for the landslide is developed. Then in Section~\ref{BoussinesqModel}, the Boussinesq model is recalled. In Section 4, solitary-wave solutions of the Peregrine system are found numerically. In Section 5, the numerical scheme for the Boussinesq system is explained and the numerical method is tested using the exact solutions of Section 4. Section 6 contains results of numerical runs for a few specific cases of bottom bathymetry, a parameter study of wave run-up in relation to the initial depth of the landslide, and a comparison with the shallow-water theory.

% *******************************************************************************
\section{The landslide model}\label{LandslideModel}
% *******************************************************************************
%
In this section we briefly present a mathematical model of underwater landslide motion. This process has to be addressed carefully since it determines the subsequent formation of water waves at the free surface. In the present study, we will assume the movable mass to be a solid body with a prescribed shape and known physical properties. Since the landslide mass and volume is preserved during the evolution, it is sufficient to determine the position of the barycenter $x = x_c(t)$ as a proxy for the motion of the whole body. As observed in the introduction, most studies of wave generation due to underwater landslides are based on \emph{prescribed bottom motion}, or on solving the equation of motion on a uniform slope while taking account of different types of friction and viscous terms. Examples of such works are \cite{DiRisio2009, Pelinovsky1996, Watts2000}. A more general approach was recently pioneered by Khakimzyanov and Shokina in \cite{KhakimzyanovG.S.Shokina2010}, where curvature effects of the bottom topography were taken into account. Since this model is applicable to a wider range of cases,
we follow the approach of \cite{KhakimzyanovG.S.Shokina2010}. However, in addition to the effects included
by Khakimzyanov and Shokina, our model also incorporates the effect of internal friction in the slide material 
(given by the dissipative force $F_i$) and the action of bottom friction, given by $F_b$.

The static bathymetry is prescribed by a sufficiently smooth single-valued function $z = -h_0(x)$, and the landslide shape is initially prescribed by a localised function $z = \zeta_0(x)$. To be specific, in this study we choose the following shape function for the landslide mass:
\begin{equation}\label{eq:cos}
  \zeta_0(x) = A\left\{
  \begin{array}{lc}
  \frac{1}{2}\Bigl(1+\cos(\frac{2\pi(x-x_0)}{\ell})\Bigr), & |x - x_0| \leq \frac{\ell}{2}\\
  0, & |x - x_0| > \frac{\ell}{2}.
  \end{array}
  \right.
\end{equation}
In this formula, $A$ is the maximum height, $\ell$ is the length of the slide and $x_0$ is the initial position of its barycenter. It is clear that the model description given below and the method of numerical integration used in the present work is applicable to any other smooth profile,
as long as it is sufficiently localized and fully submerged.

Since the landslide motion is translational, its shape at time $t$ is given by the function $z = \zeta(x,t) = \zeta_0(x-x_c(t))$. Recall that the landslide center is located at a point with abscissa $x = x_c(t)$. Then, the impermeable bottom for the water wave problem can be easily determined at any time by simply superposing the static and dynamic components. Thus the bottom boundary conditions for the fluid are to be imposed at
\begin{equation*}
  z = -h(x,t) = -h_0(x) + \zeta(x,t).
\end{equation*}

To simplify the subsequent presentation, we introduce the classical arc-length parameterisation, where the parameter $s = s(x)$ is given by the formula
\begin{equation}\label{eq:len}
  s = L(x) = \int_{x_0}^{x}\sqrt{1+(h_0'(\xi))^2}\,\ud\xi.
\end{equation}
The function $L(x)$ is monotone and can be efficiently inverted to yield the original Cartesian abscissa $x = L^{-1}(s)$. Within the parameterisation \eqref{eq:len}, the center of the landslide is initially located at a point with the curvilinear coordinate $s = 0$. The local tangential direction is denoted by $\tau$ and the normal direction by $n$.

A straightforward application of Newton's second law reveals that the landslide motion is governed by the differential equation
\begin{equation*}
  m\odd{s}{t} = F_\tau(t),
\end{equation*}
where $m$ is the landslide mass and $F_\tau(t)$ is the tangential component of the sum of forces acting on the moving submerged body. In order to project the forces onto the axes of the local coordinate system, the angle $\theta(x)$ between $\tau$ and $Ox$ is needed. This angle is determined by
\begin{equation*}
  \theta(x) = -\arctan\bigl(h'_0(x)\bigr).
\end{equation*}

Let us denote by $\rho_w$ and $\rho_\ell$ the densities of the water and landslide material correspondingly. If $V$ is the volume of the slide, then the total mass $m$ is given by the expression
\begin{equation}\label{genMass}
  m := \bigl(\rho_\ell + c_w\rho_w\bigr)V,
\end{equation}
where $c_w$ is the added mass coefficient. As explained in \cite{Batchelor2000}, a portion of the water mass has to be added to the mass of the landslide since it is entrained by the underwater body motion. For a cylinder, the coefficient $c_w$ is equal exactly to one, but in the present case, the coefficient has to be estimated. The volume of the sliding material is given by $ V = W\cdot S$, where $W$ is the landslide width in the transverse direction, and $S$ can be computed by
\begin{equation*}
S = \int_{\R}\zeta_0(x) \,\ud x.
\end{equation*}
The last integral can be computed exactly for the particular choice \eqref{eq:cos} of the landslide shape to give
\begin{equation*}
  V = \frac12\ell A W.
\end{equation*}

The total projected force acting on the landslide can be conventionally represented as a sum of the force $F_g$ representing the joint action of gravity and buoyancy, and the total contribution of various dissipative forces.

\begin{figure}
\centering
\includegraphics[width=0.69\textwidth]{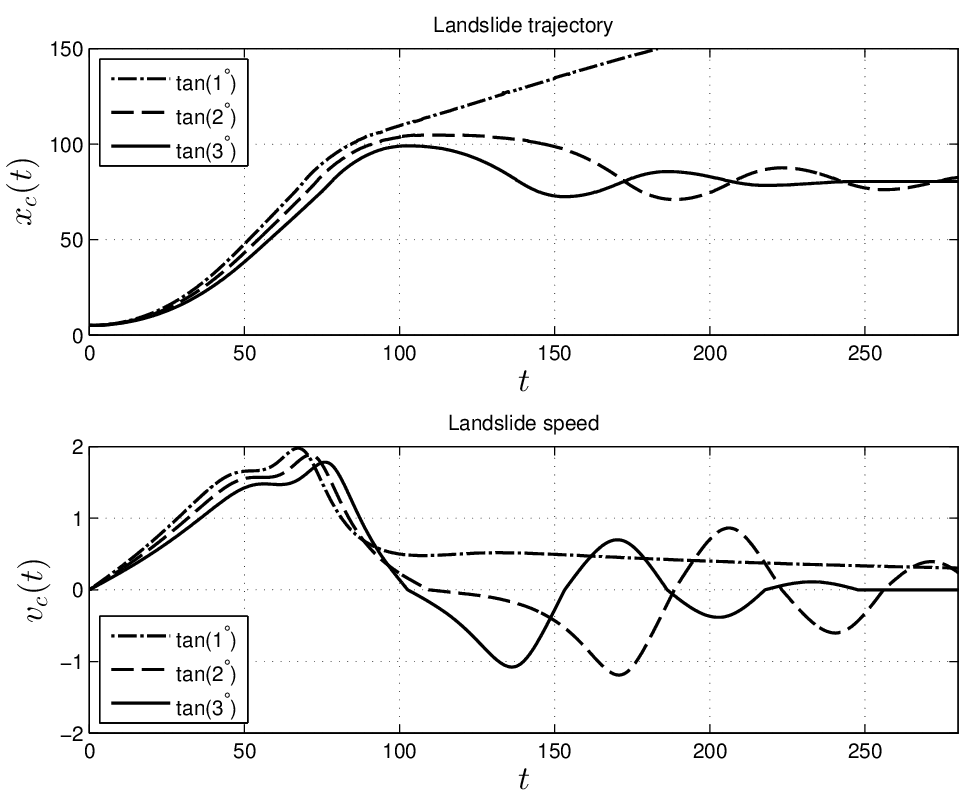}
\caption{\small\em Position and velocity of the barycenter of the landslide as functions of dimensional time for three different values of the friction coefficient $c_f$.}
\label{SlideSpeed}
\end{figure}
\begin{figure}
  \centering
  \includegraphics[width=0.69\textwidth]{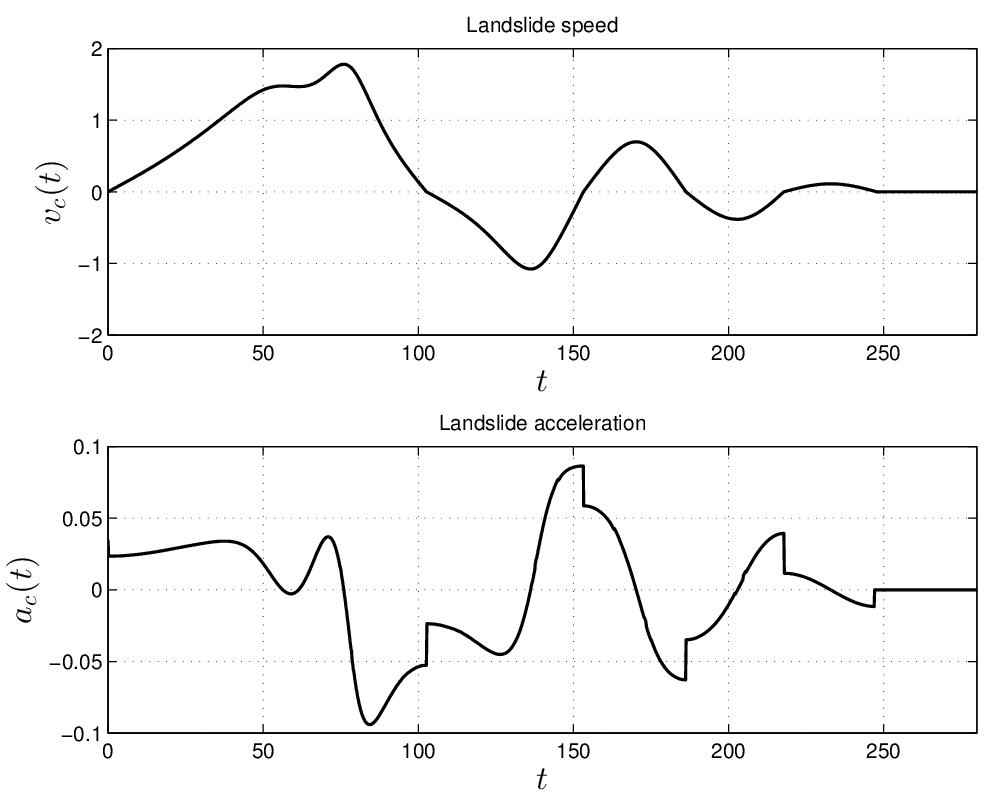}
  \caption{\small\em Velocity and acceleration of the barycenter of the landslide as a function of dimensional time. The friction coefficient is $c_f = \tan(3^{\circ})$. The discontinuities in the acceleration are due to the coefficient $\sign\bigl(\od{s}{t}\bigr)$ in the definition of the friction force.}
 \label{SpeedAccel}
\end{figure}

The gravity and buoyancy forces act in opposite directions and their horizontal 
projection $F_g$ can be easily computed by
\begin{equation*}
  F_g(t) = (\rho_\ell - \rho_w)W g\int_\R \zeta(x,t)\sin\bigl(\theta(x)\bigr)\,\ud x.
\end{equation*}
Now, let us specify the dissipative forces. The water resistance to the motion of the landslide $F_r$ due to viscous dissipation is proportional to the maximal transverse section of the moving body and to the square of its velocity. In addition, the coefficient $\sign\bigl(\od{s}{t}\bigr)$ is needed to dissipate the landslide kinetic energy independently of its direction of motion. Thus the force $F_r$ takes the form
\begin{equation*}
  F_r = - \sign \bigg( \od{s}{t} \bigg) \frac12 c_d\rho_w AW\Bigl(\od{s}{t}\Bigr)^2,
\end{equation*}
where $c_d$ is the resistance coefficient of the water. The friction force $F_f$ is proportional to the normal force exerted on the body due to the weight:
\begin{equation*}
  F_f = -c_f \sign \bigg( \od{s}{t} \bigg) N(x,t).
\end{equation*}
The normal force $N(x,t)$ is composed of the normal components of gravity and buoyancy forces, but also of the centripetal force due to the variation of the bottom slope:
\begin{equation*}
  N(x,t) = (\rho_\ell - \rho_w) g W\int_\R\zeta(x,t)\cos\bigl(\theta(x)\bigr)\,\ud x 
         + \rho_\ell W\int_\R\zeta(x,t)\kappa(x)\Bigl(\od{s}{t}\Bigr)^2\,\ud x.
\end{equation*}
Here $\kappa(x)$ is the signed curvature of the bottom which can be computed using the formula
\begin{equation*}
  \kappa(x) = \frac{h''_0(x)}{\bigl(1+(h'_0(x))^2\bigr)^{\frac32}}.
\end{equation*}
We note that the last term vanishes for a plane bottom since $\kappa(x) \equiv 0$ in this particular case. Energy loss inside the sliding material due to internal friction is modeled by 
\begin{equation*}
  F_i = -c_v \rho_\ell W S \, \od{s}{t},
\end{equation*}
where $c_v$ is an internal friction coefficient. Finally, dissipation in the boundary layer between the landslide and the solid bottom is taken account of by the term
\begin{equation*}
  F_b = - c_b \rho_w W \ell \, \od{s}{t} \bigg|\od{s}{t}\bigg|,
\end{equation*}
where $c_b$ is the Ch\'{e}zy coefficient.

Finally, if we sum up the contributions of all the forces described above, we obtain the second order differential equation
\begin{multline}\label{eq:ODE}
  (\gamma + c_w)S\odd{s}{t} = (\gamma-1)g \Bigl(\It_1(t) - c_f\sigma(t)\It_2(t)\Bigr) \\ -\sigma(t)\Bigl(c_f\gamma\It_3(t) + \frac12 c_d A\Bigr)\Bigl(\od{s}{t}\Bigr)^2
  - c_v\gamma S\od{s}{t} - c_b\ell\od{s}{t}\left|\od{s}{t}\right|,
\end{multline}
where $\gamma := \frac{\rho_\ell}{\rho_w} > 1$ is the ratio of densities, $\sigma(t) := \sign\Bigl(\od{s}{t}\Bigr)$ and the integrals $\It_{1,2,3}(t)$ are defined by
\begin{equation*}
  \It_1(t) = \int_\R\zeta(x,t)\sin\bigl(\theta(x)\bigr)\,\ud x,
\end{equation*}
\begin{equation*}
  \It_2(t) = \int_\R\zeta(x,t)\cos\bigl(\theta(x)\bigr)\,\ud x,
\end{equation*}
\begin{equation*}
  \It_3(t) = \int_\R \zeta(x,t)\kappa(x)\,\ud x.
\end{equation*}
In order to obtain a well-posed initial value problem, equation \eqref{eq:ODE} has to be supplemented with initial conditions for $s(0)$ and $s'(0)$. In the remainder we always take homogeneous initial conditions, and consider the motion driven only by the gravitational acceleration of the landslide. However, different boundary conditions might also be reasonable from a modeling point of view.
\begin{figure}
\centering
\includegraphics[width=0.62\textwidth]{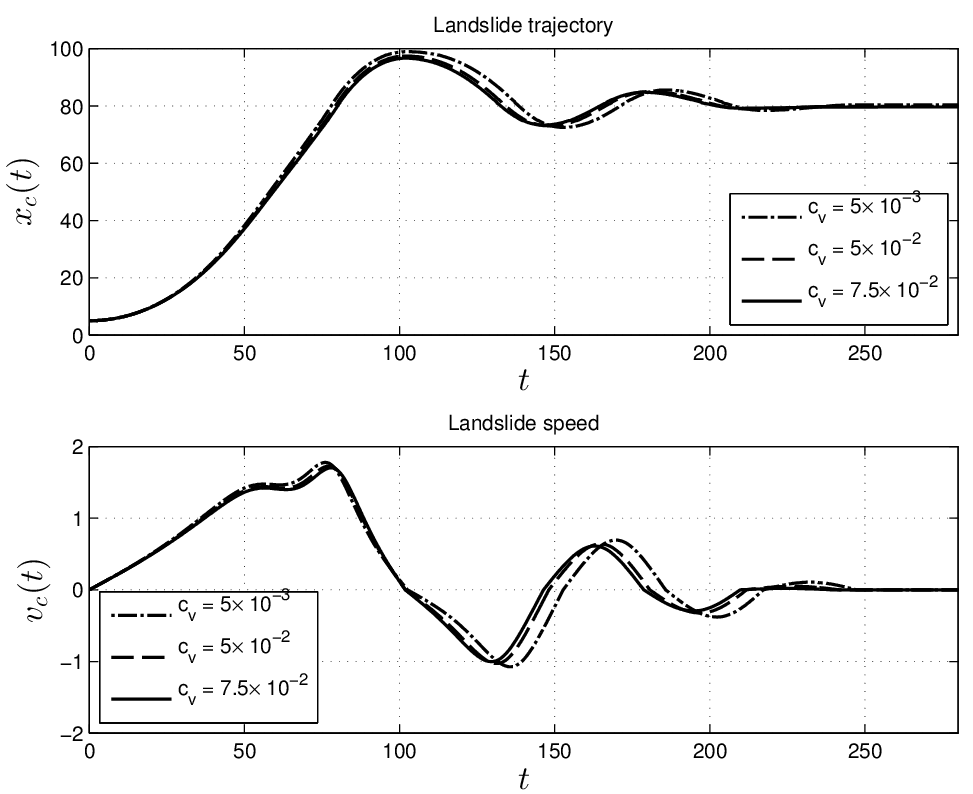}
\caption{\small\em Position and velocity of the barycenter of the landslide as functions of dimensional time for three different values of the friction coefficient $c_v$.}
\label{fig:cv}
\end{figure}
\begin{figure}
\centering
\includegraphics[width=0.62\textwidth]{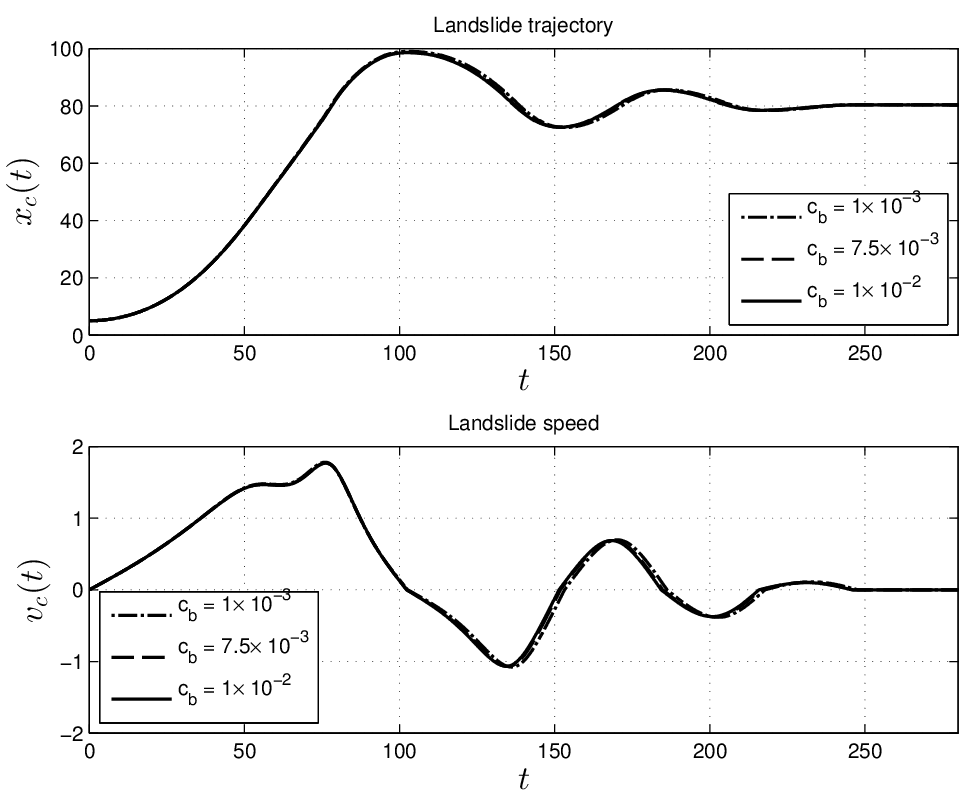}
\caption{\small\em Position and velocity of the barycenter of the landslide as functions of dimensional time for three different values of the friction coefficient $c_b$.}
\label{fig:cb}
\end{figure}

In order to approximate solutions of equation \eqref{eq:ODE}, we employ the Bogacki-Shampine third-order Runge-Kutta scheme. The integrals $\It_{1,2,3}(t)$ are computed using the trapezoidal rule, and once the landslide trajectory $s = s(t)$ is found, we use equation \eqref{eq:len} to find its motion $x = x_c(t)$ in the initial Cartesian coordinate system. This yields the bottom motion that drives the fluid solver.

For illustrative purposes we show a few examples of landslide trajectories over the bottom profile depicted in Figure~\ref{SlideBathy}. The other parameters used in the simulations are given in Section~\ref{Results} and also in Table~\ref{tab:pars}. We performed a series of simulations in order to study the effect of various dissipative terms on the landslide trajectory. The dependence on the friction coefficient $c_f$ is shown in Figure~\ref{SlideSpeed} where the landslide barycenter position $x_c(t)$ and its velocity $v_c(t)$ are shown as functions of time for $c_f = \tan(1^\circ)$, $\tan(2^\circ)$ and $\tan(3^\circ)$. Ìn the case of the weak friction $c_f = \tan(1^\circ)$, the landslide reaches a sufficient speed to escape from the basin depicted on Figure~\ref{SlideBathy}. For the latter case ($c_f = \tan(3^\circ)$) we show also simultaneously the landslide speed $v_c(t) := \od{x_x}{t}$ and its acceleration $a_c(t) := \od{v_c}{t} = \od{^2 x_c}{t^2}$ in Figure \ref{SpeedAccel}. In particular, one can see that the acceleration is a discontinuous function whose jumps correspond exactly to moments of time where the speed $v_c$ changes its sign, in accordance with the employed model \eqref{eq:ODE}. However, in our model there are also two new dissipative terms $F_i$ and $F_b$ whose importance has to be studied also. 
We fix the value of $c_f = \tan(3^\circ)$ for all subsequent experiments and we will vary only the two other coefficients $c_v$ and $c_b$ for fixed other parameters given in Table~\ref{tab:pars}. These numerical results are presented in Figures~\ref{fig:cv} and \ref{fig:cb}. One can see that the influence of these parameters on the landslide trajectory is weaker. However, we choose to keep them in the model in order to have more latitude for fine tuning the slide trajectory if need be.

%
% **********************************************************************************
\section{The Boussinesq model}\label{BoussinesqModel}
% **********************************************************************************
%
Once the motion of the landslide is determined, and therefore the time-dependent
bathymetry $h(x,t) = h_0(x) - \zeta(x,t)$ is given, the next step is to consider the coupling between the bathymetry variations and the evolution of surface waves.
The main assumptions on the fluid are that it is inviscid and incompressible, and that the flow is irrotational. Under these assumptions, the potential-flow free surface problem governs the motion of the fluid. However, in the present case, the fluid is shallow, and the waves at the surface are of small amplitude when compared to the depth of the fluid. In that case, the potential-flow problem may be simplified, and the model used in this paper is a variant of the so-called classical Boussinesq system derived by Boussinesq \cite{bouss}. 

Let us first consider the case of an even bottom, and a constant fluid depth $d_0$. Denote a typical wave amplitude  by $a$, and a typical wavelength by $\lambda$. The parameter $\alpha = \frac{a}{d_0}$ then describes the relative amplitude of the waves, and the parameter $\beta = \frac{d_0^2}{\lambda^2}$ measures the 'shallowness' of the fluid in comparison to the wavelength. In the case when both $\alpha$ and $\beta$ are small and approximately of the same order of magnitude, the system
\begin{align}\label{Peregrine}
\begin{split}
\eta_t + d_0 u_x + (\eta u)_x = 0,\\
 u_t + g \eta_x+ uu_x - \frac{d_0^2}{3} u_{xxt} = 0
\end{split}
\end{align}
may be used as an approximate model for the description of the evolution of the surface waves and the fluid flow. In \eqref{Peregrine}, $\eta$ denotes the deflection of the free surface from its rest position, and $u$ denotes the horizontal fluid velocity at a height $z=d_0(-1+ \sqrt{1/3} )$ in the fluid column if $z$ is measured from the rest position of the free surface. The same equation appears if the velocity is taken to be the average of the horizontal velocity over the flow depth.

The system \eqref{Peregrine} was first derived by Peregrine in \cite{Peregrine1967},
and falls into a general class of Boussinesq systems, as shown in the systematic
studies \cite{BCS, Nwogu1993}. As opposed to the shallow-water approximation, the pressure is not assumed to be hydrostatic, and the horizontal velocity varies with depth. In fact, the horizontal velocity profile is a quadratic function of $z$ \cite{Whitham1999}. Non-hysrostatic effects lead the appearance of linear dispersive terms in the governing equations. The problem of landslide generated waves has been addressed in the fully nonlinear shallow water framework \cite{Watts2003, Chubarov2005, Shokin2007, Beisel2010}. Nevertheless, several authors obtained recently interesting results even in the linear \cite{Sammarco2008, Seo2013} or nonlinear \cite{Fernandez-Nieto2007, Didenkulova2010, Beisel2012} hydrostatic models.

The derivation of \eqref{Peregrine} given in \cite{Peregrine1967} also featured and extension to non-constant but time-independent bathymetry. However, the present case of a dynamic bottom profile calls for a system which allows for time-dependent bathymetry, and such a system was derived in \cite{Wu1987}. Given a bottom topography described by $z = -h(x,t)$, the system takes the form
\begin{align}\label{PeWu}
\begin{split}
\eta_t + \bigl((h+\eta)u\bigr)_x + h_t = 0,\\
 u_t + g\eta_x+ uu_x = \frac{1}{2} h \bigl(h_t + (hu)_x\bigr)_{xt} - \frac{h^2}{6}u_{xxt}.
\end{split}
\end{align}
In order for this system to be asymptotically valid, we need $\alpha \sim \beta$ as before. Moreover, concerning the unsteady bottom profile, we make the assumptions that $h_x \le \O(\alpha \beta^{1/2})$, and $h_t \le \O(\alpha \beta^{1/2})$.

In comparison to the shallow-water equations with a time-dependent bottom topography, 
the system \eqref{PeWu} has additional terms on the right-hand side of the second equation. The effect of these terms is to incorporate frequency dispersion into the model. One practical aspect of this modification is that wave breaking can be completely avoided as long as the amplitude of the waves is small enough. Wave breaking is also possible in evolution systems of Boussinesq type \cite{Bjorkavag2011,Briganti}, but the amplitudes occurring in the present problem are far from the breaking limit. The phase speed of a small-amplitude linear wave of wavelength $2\pi / k$ in the equation \eqref{PeWu} 
with a stationary even bottom has the form
$$c^2 = \frac{g d_0}{1+\frac{d_0^2}{3}k^2 },$$ while the phase speed is given by
$$c^2 = g d_0 \frac{\tanh (kd_0)}{kd_0}$$ in the linearized full water wave problem.
Thus one might argue that the dispersion in \eqref{PeWu} is too strong in comparison 
with dispersion in realistic water waves. 
However, as discussed in \cite{Bjorkavag2011}, 
the linear dispersion relation of \eqref{PeWu} is still closer to the dispersion relation 
of the original water-wave problem than most other standard Boussinesq equations 
which feature even faster decay of the phase speed with increasing $k$.

% **********************************************************************************
\section{Solitary waves}\label{SolitaryWaves}
% **********************************************************************************

Before the numerical method for approximating solutions of \eqref{PeWu} is presented, we digress for a moment, and explain how to find numerically exact solutions of the system \eqref{Peregrine}. These solutions will later be used to test the implementation of the numerical procedure. Assuming the special form
\begin{equation*}
  \eta(x,t) = \eta(\xi), \quad u(x,t) = u(\xi), \quad \xi := x - c_s t,
\end{equation*}
and substituting this representation into the governing equations \eqref{Peregrine}, there appears
\begin{eqnarray*}
  -c_s\eta' + \bigl((d+\eta)u\bigr)' &=& 0, \\
  -c_s u' + \frac{1}{2}(u^2)' + g\eta' + c_s\frac{d^2}{3}u''' &=& 0.
\end{eqnarray*}
Assuming decay of both $\eta$ and $u$ to zero as $|x| \rightarrow \infty$, the integration of the mass conservation equation from $-\infty$ to $\xi$ gives the following relation between $\eta$ and $u$:
\begin{equation}\label{eq:etau}
  u = \frac{c_s\eta}{d+\eta}, \qquad \eta = \frac{d\cdot u}{c_s-u}.
\end{equation}
The momentum balance equation can now be integrated to yield
\begin{equation}\label{eq:cPerSW}
  -c_s\Big(u - \frac{d^2}{3}u''\Big) + \frac{1}{2} u^2 + g\eta  = 0.
\end{equation}

\begin{figure}
\centering
 \subfigure[]{\includegraphics[width=0.49\textwidth]{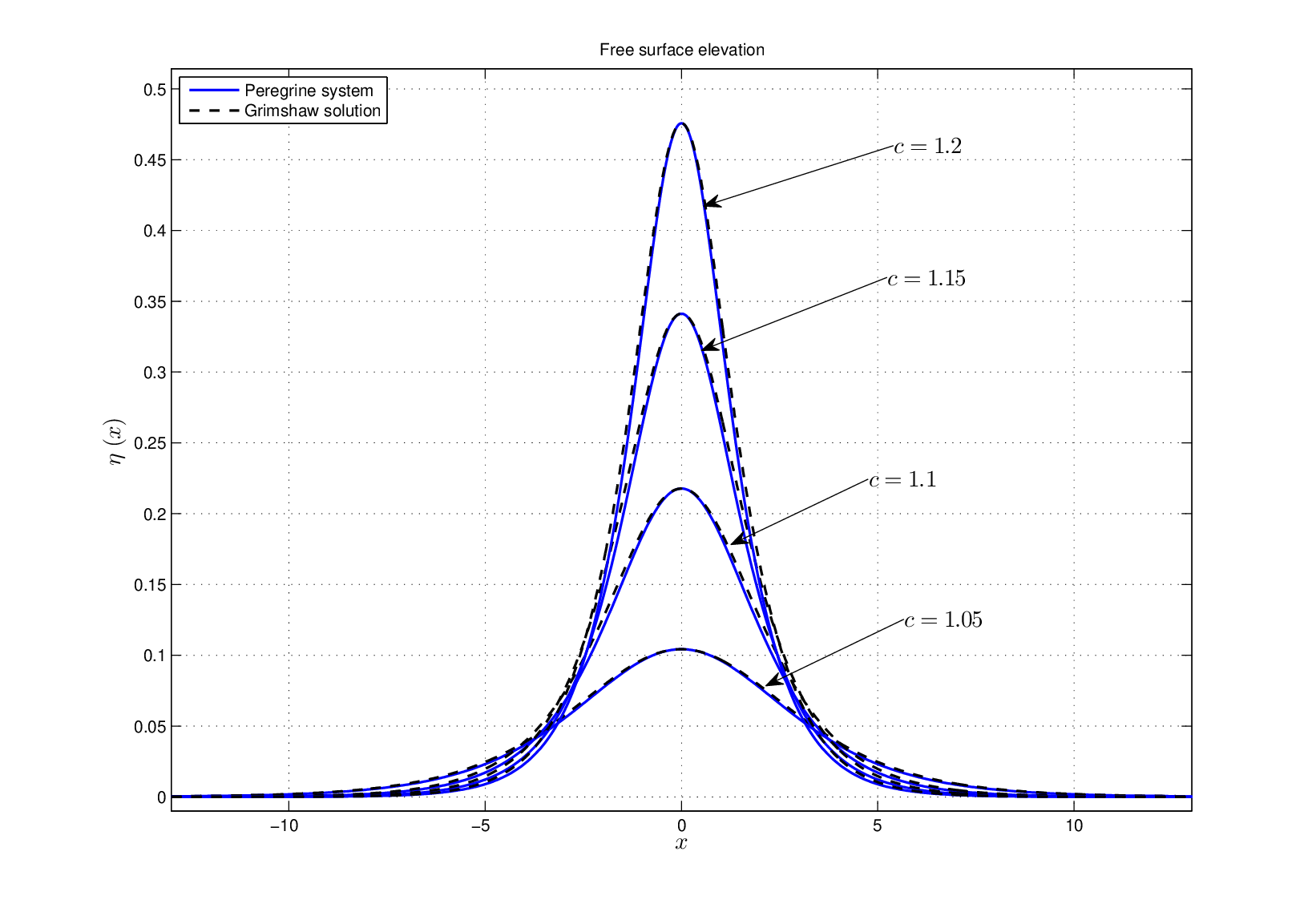}}
 \subfigure[]{\includegraphics[width=0.49\textwidth]{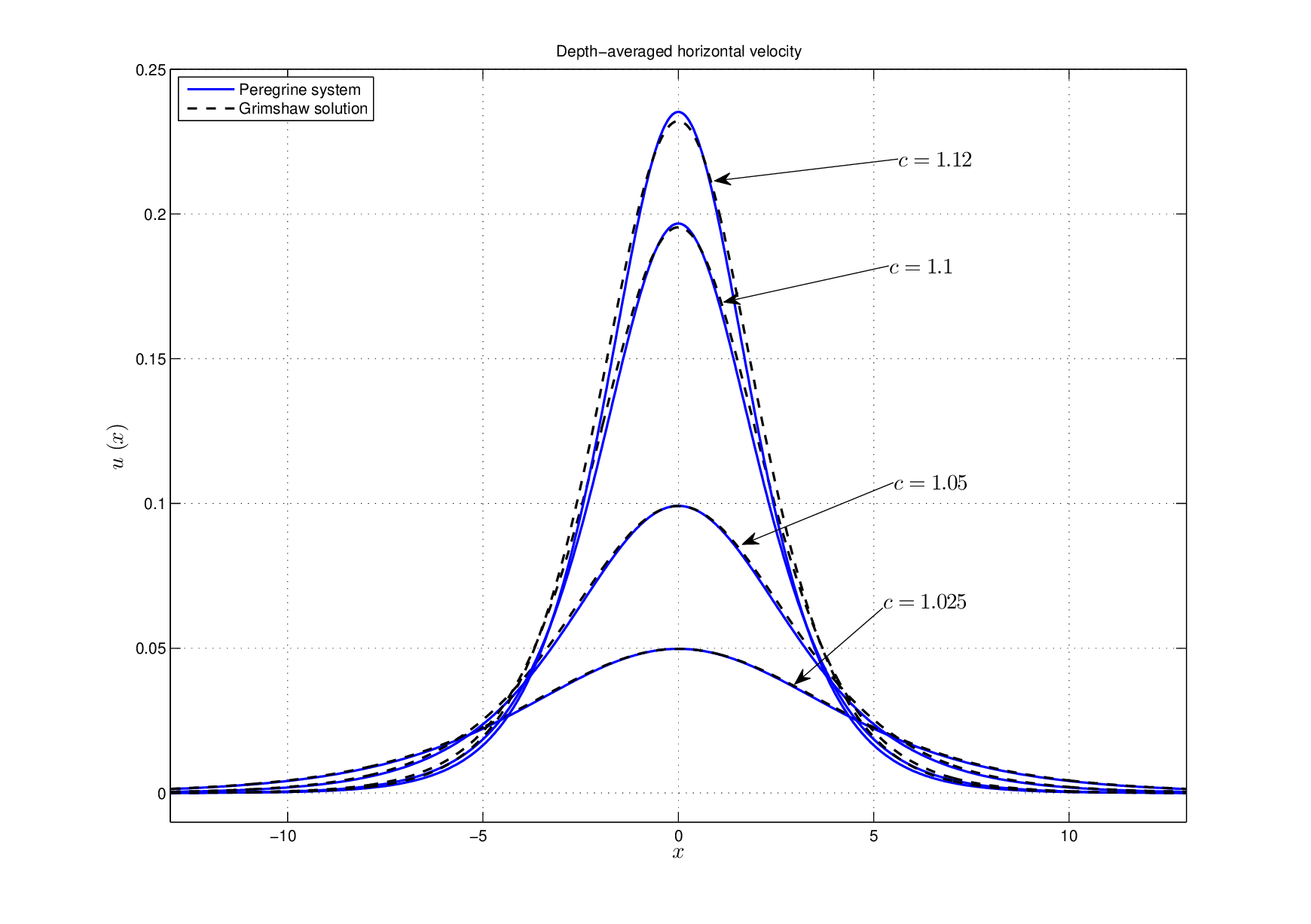}}
\caption{\small\em Comparison of the numerical approximation of solitary wave solutions of \eqref{Peregrine} to Grimshaw's third-order asymptotic approximation of solitary waves using the Euler equations for the full water wave problem. The left panel shows the surface elevation, and the right panel shows the horizontal velocity at $z = d_0(-1+ \sqrt{1/3})$.}
\label{Grimshaw}
\end{figure}

Finally, in order to obtain a closed form equation in terms of the velocity $u$, 
we substitute the expression \eqref{eq:etau} for $\eta$ into \eqref{eq:cPerSW}.
The resulting differential equation can be written in operator notation as
\begin{equation*}
  \LL u = \N(u),
\end{equation*}
where the linear operator $\LL$ and the nonlinear operator $\N$, are defined respectively by
\begin{equation*}
\LL u := c_s\Big(u - \frac{d^2}{3}u''\Big), \quad \mbox{ and }
  \N(u) := \frac{1}{2} u^2 + \frac{gdu}{c_s-u}.
\end{equation*}
While nothing formal appears to be known about existence of localised solutions
of \eqref{eq:etau}, \eqref{eq:cPerSW}, it is straightforward to compute approximations of solitary waves numerically. In particular, one may use the well known Petviashvili iteration method which takes the form
\begin{equation}\label{eq:perv}
  u_{n+1} = \LL^{-1}\scal\N(u_n)\scal\Biggl(\frac{(u_n, \N(u_n))}{(u_n, \LL u_n)}\Biggr)^{-q}.
\end{equation}
The exponent $q$ is usually defined as a function of the degree $p$ of the nonlinearity, with the rule of thumb that the expression $q  := \frac{p}{p-1}$ generally works well. In our case, the nonlinearities are quadratic, so that we choose $p = 2$, and hence $q = 1$.  

\begin{figure}
\centering
\includegraphics[width=0.69\textwidth]{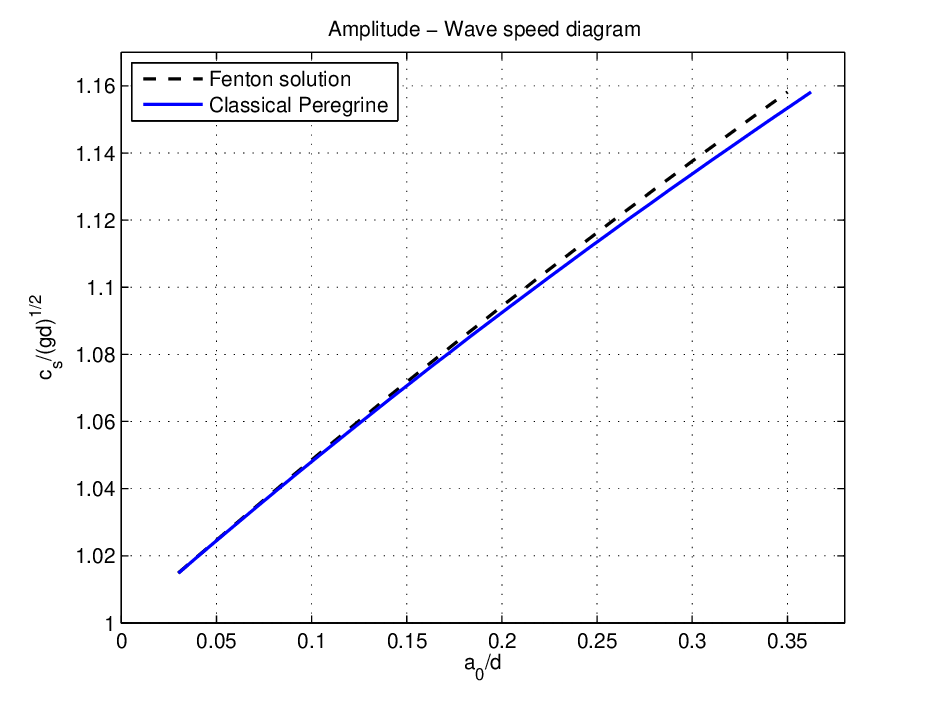}
\caption{\small\em Amplitude-speed relation of solitary wave solutions of \eqref{Peregrine} and of Fenton's ninth-order asymptotic approximation of solitary waves using the Euler equations for the full water wave problem.}
\label{AmpSpeed}
\end{figure}

The Petviashvili method was analyzed in \cite{DPelinovsky2004}, and can be very efficiently implemented using the Fast Fourier Transform \cite{Frigo2005}. The iteration can be started for instance with the third-order asymptotic solution of Grimshaw \cite{Grimshaw1971}. The iterative procedure is continued until the $L_\infty$ norm between two successive iteration is on the order of machine precision. Figure~\ref{Grimshaw} shows approximate solitary-wave solutions of \eqref{Peregrine} with various wave speeds, and compares them to the third-order asymptotic approximation of solitary-wave solutions of the full water-wave problem obtained by Grimshaw \cite{Grimshaw1971}. The left panel shows comparisons of the
free-surface excursion, while the right panel shows a comparison of the horizontal component of the velocity field, evaluated at the non-dimensional height $\tilde{z}$ given by $\tilde{z} = -1 + \sqrt{1/3}$. Figure~\ref{AmpSpeed} shows a comparison of the wavespeed-amplitude relation between the solitary-wave approximation of \eqref{Peregrine} and the ninth-order asymptotic approximation to the full water-wave problem obtained by Fenton \cite{Fenton1972}.

% **********************************************************************************
\section{The numerical scheme}\label{NumericalScheme}
% **********************************************************************************

For the numerical discretisation, a finite-volume discretisation procedure similar to the one used in \cite{Barth1994, Barth2004} is employed. Let us take as a unit of length the undisturbed depth $d_0$ of the fluid above the barycenter of the landslide, and as a unit of time the ratio $\sqrt{\frac{d_0}{g}}$. Then the Peregrine system \eqref{PeWu} is rewritten in terms of the total water depth $H$ as
\begin{align}
  H_t\ +\ [Hu]_x\ = &\ 0, \label{eq:per1} \\
  u_t\ +\ \bigl[\half u^2 + (H-h)\bigr]_x\ =
              & \ \frac12 hh_{xtt} + \frac12 h(hu)_{xxt} - \frac16 h^2 u_{xxt},
 \label{eq:per2}
\end{align}
The system \eqref{eq:per1}, \eqref{eq:per2} can be formally rewritten in the form
\begin{equation}\label{eq:conslaw}
  \V_t\ +\ [\,\F(\V)\,]_x\ =\ \S_b \ + \ \M(\V),
\end{equation}
where the density $\V$ and the advective flux $\F(\V)$ are defined by
\begin{equation*}
  \V\ \equiv\ \begin{pmatrix}
        H \\
        u
      \end{pmatrix}, \quad
  \F(\V)\ \equiv\ \begin{pmatrix}
           H\,u \\
           \half\,u^2\, +\, (H-h)
         \end{pmatrix}.
\end{equation*}
The source term is defined by 
\begin{equation*}
  \S_b \ \equiv\ \begin{pmatrix}
        0 \\
        \frac12 h h_{xtt}
      \end{pmatrix}, 
\end{equation*}
and the dispersive term is defined by
\begin{equation*}
  \M(\V)\ \equiv\ \begin{pmatrix}
        0 \\
         \frac12 h(hu)_{xxt} - \frac16 h^2 u_{xxt} 
      \end{pmatrix}.
\end{equation*}
We begin our presentation by a discretisation of the hyperbolic part of \eqref{eq:per1}, \eqref{eq:per2}, which is the classical nonlinear shallow-water system, and then discuss the treatment of dispersive terms. The Jacobian of the advective flux $\F(\V)$ is easily computed to be
\begin{equation*}
  \A(\V)\ =\ \pd{\F(\V)}{\V}\ =\  
  \begin{pmatrix}
    u & H \\
    1 & u
  \end{pmatrix},
\end{equation*}
and it is clear that $\A(\V)$ has the two distinct eigenvalues
\begin{equation*}
  \lambda^\pm\ =\ u\ \pm\ c_s, \qquad c_s\ \equiv\ \sqrt{H}.
\end{equation*}
The corresponding right and left eigenvectors are the columns of the matrices
\begin{equation*}
  R\ =\ \begin{pmatrix}
        H & -H \\
        c_s & c_s
  	  \end{pmatrix}, \qquad
  L\ =\ R^{-1}\ =\ \frac12
  		      \begin{pmatrix}
                 H^{-1} & c_s^{-1} \\
                 -H^{-1} & c_s^{-1}
               \end{pmatrix}.
\end{equation*}

We consider a partition of the real line $\R$ into cells (or finite volumes) $\C_i = [x_{i-\frac12}, x_{i+\frac12}]$ with cell centers $x_i = \half(x_{i-\frac12} + x_{i+\frac12})$ ($i\in \Z$). Let $\Delta x_i$ denote the length of the cell $\C_i$. In the sequel we will consider only uniform partitions with $\Delta x_i$ = $\Delta x$, $\forall i\in\Z$. We would like to approximate the solution $\V(x,t)$ by discrete values. In order to do so, we introduce the cell average of $\V$ on the cell $\C_i$ (denoted with an overbar), i.e., 
\begin{equation*}
\bar{\V}_i(t)\ \equiv\ \bigl(\,\hm_i(t)\,,\,\um_i(t)\,\bigr)\ =\ 
\frac{1}{\Delta x} \int_{\C_i} \V(x,t)\,\ud\/x.
\end{equation*}
A simple integration of \eqref{eq:conslaw} over the cell $\C_i$ leads to the exact relation
\begin{equation*}
  \od{\bar{\V}}{t}\ +\ \frac{1}{\Delta x}\Bigl[\,\F(\V(x_{i+\frac12},t))
  \, -\, \F(\V(x_{i-\frac12},t))\,\Bigr]\ =\  
  \frac{1}{\Delta x}\int_{\C_i}\S_b(\V)\,\ud\/x\ \equiv\ \bar{\S}_i.
\end{equation*}
Since the discrete solution is discontinuous at cell interfaces $x_{i+\frac12}$ ($i\in\Z$), we replace the flux at the cell faces by the so-called numerical flux function
\begin{equation*}
  \F(\V(x_{i\pm\frac12},t))\ \approx\ \F_{i\pm\frac12}(
  \bar{\V}_{i\pm\frac12}^{L},\bar{\V}_{i\pm\frac12}^{R}),
\end{equation*}
where $\bar{\V}_{i\pm\frac12}^{L,R}$ denotes the reconstructions of the conservative variables $\bar{\V}$ from left and right sides of each cell interface (the reconstruction procedure employed in the present study will be described below). Consequently, the semi-discrete scheme takes the form
\begin{equation}\label{eq:si1}
\od{\bar{\V}_i}{t}\ +\ \frac{1}{\Delta x}\bigl[\,\F_{i+\frac12}\, -\, 
\F_{i-\frac12}\,\bigr]\ =\ \bar{\S}_i.
\end{equation}

In order to discretise the advective flux $\F(\V)$, we follow the method of \cite{Ghidaglia1996, Ghidaglia2001} and use the following FVCF scheme 
\begin{equation*}
  \F(\V,\W)\ =\ \frac{\F(\V)\,+\,\F(\W)}{2}\ -\ \U(\V,\W)\scal \frac{\F(\W)\,-\,\F(\V)}{2}.
\end{equation*}
The first part of the numerical flux is centered, the second part is the upwinding introduced through the Jacobian sign-matrix $\U(\V,\W)$ defined by
\begin{equation*}
  \U(\V,\W)\ =\ \sign\bigl[\A(\half(\V + \W))\bigr], \qquad 
  \sign(\A)\ =\ R \scal\diag(s^+, s^-)\scal L,
\end{equation*}
where $s^\pm \equiv \sign(\lambda^\pm)$. After some simple algebraic computations, one can find
\begin{equation*}
  \U\ =\ \frac{1}{2}\begin{pmatrix}
   		s^+ + s^- & (H/c_s)\,(s^+ - s^-) \\
   		(c_s/H)\,(s^+ - s^-) & s^+ + s^-
      \end{pmatrix},
\end{equation*}
the sign-matrix $\U$ being evaluated at the average state of left and right values.

Finally the source term $\S_b(x,t) = (0, \half hh_{xtt})$, which is due to the moving bottom, is discretised by evaluating the bathymetry function and its derivatives at cell centers:
\begin{equation*}
  \frac{1}{\Delta x}\int_{\C_i}\S_b(x,t)\,\ud\/x 
             \approx \big( 0,{\textstyle \frac12} h(x_i,t)\,h_{xtt}(x_i,t) \big).
\end{equation*}
Recall that the bathymetry is composed of the static part and of the landslide subject to a translational motion:
\begin{equation*}
  h(x,t) = h_0(x) - \zeta(x,t) = h_0(x) - \zeta_0 \bigl(x - x_c(t)\bigr).
\end{equation*}
The derivative $h_{xtt}$ can be readily obtained from the formula
\begin{equation*}
  h_{xtt}(x,t) = \odd{ x_c}{t}\odd{\zeta_0}{x}(x-x_c(t))
  - \Bigl(\od{x_c}{t}\Bigr)^2 \oddd{\zeta_0}{x}(x-x_c(t)).
\end{equation*}

% **********************************************************************************
\subsection{High-order reconstruction}
% **********************************************************************************

In order to obtain a higher-order scheme in space, we need to replace the piecewise constant data by a piecewise polynomial representation. This goal is achieved by various so-called reconstruction procedures such as MUSCL TVD \cite{Kolgan1975, Leer1979, Leer2006}, UNO \cite{HaOs}, ENO \cite{Harten1989}, WENO \cite{Xing2005} and many others. In recent studies on unidirectional wave models \cite{Dutykh2010e} and on Boussinesq-type equations \cite{Dutykh2010}, the UNO2 scheme showed a good performance with small dissipation in realistic propagation and run-up simulations. Consequently, we retain this scheme for the discretisation of the advective flux of the Peregrine system \eqref{eq:per1}, \eqref{eq:per2}.

The main idea of the UNO2 scheme is to construct a non-oscillatory piecewise-parabolic interpolant $\Q(x)$ to a piecewise smooth function $\V(x)$ (see \cite{HaOs} for more details). On each segment containing the face $x_{i+\frac12} \in [x_i, x_{i+1}]$, the function $\Q(x) = \q_{i+\frac12}(x)$ is locally a quadratic polynomial and wherever $v(x)$ is smooth we have
\begin{equation*}
  \Q(x)\ -\ \V(x)\ =\ \boldsymbol{0}\ +\ \O(\Delta x^3), \qquad
  \od{\Q}{x}(x\pm 0)\ -\ \od{\V}{x}\ =\ \boldsymbol{0}\ +\ \O(\Delta x^2).
\end{equation*}
Also, $\Q(x)$ should be non-oscillatory in the sense that the number of its local extrema does not exceed that of $\V(x)$. Since $\q_{i+\frac12}(x_i) = \bar{\V}_i$ and $\q_{i+\frac12}(x_{i+1}) = \bar{\V}_{i+1}$, it can be written in the form
\begin{equation*}
  \q_{i+\frac12}(x)\ =\ \bar{\V}_i\ +\ \mathfrak{d}_{i+\frac12}\{\V\}\times\frac{x - x_i}{\Delta x}\ +\  
  \half\,\mathfrak{D}_{i+\frac12}\{\V\}\times\frac{(x-x_i)(x-x_{i+1})}{\Delta x^2},
\end{equation*}
where $\mathfrak{d}_{i+\frac12} \{\V\} \equiv \bar{\V}_{i+1} - \bar{\V}_{i}$ and $\mathfrak{D}_{i+\frac12} \V$ is closely related to the second derivative of the interpolant since $\mathfrak{D}_{i+\frac12}\{\V\} = \Delta x^2\, \q''_{i+\frac12}(x)$. The polynomial $\q_{i+\frac12}(x)$ is chosen to be the least oscillatory between two candidates interpolating $\V(x)$ at $(x_{i-1}, x_i, x_{i+1})$ and $(x_i, x_{i+1}, x_{i+2})$. This requirement leads to the following choice of $\mathfrak{D}_{i+\frac12} \{\V\} \equiv \minmod\bigl(\mathfrak{D}_i \{\V\},\mathfrak{D}_{i+1}\{\V\}\bigr)$ with
\begin{equation*}
  \mathfrak{D}_i\{\V\}\ =\ \bar{\V}_{i+1}\ -\ 2\,
  \bar{\V}_i\ +\ \bar{\V}_{i-1}, \qquad
  \mathfrak{D}_{i+1}\{\V\}\ =\ \bar{\V}_{i+2}\ -\ 2\,\bar{\V}_{i+1}
  \ +\ \bar{\V}_i,
\end{equation*}
and where $\minmod(x,y)$ is the usual minmod function defined as
\begin{equation*}
  \minmod (x,y)\ \equiv\ \half\,[\,\sign(x)\,+\,\sign(y)\,]\times\min(|x|,|y|).
\end{equation*}

To achieve the second order $\O(\Delta x^2)$ accuracy, it is sufficient to consider piecewise linear reconstructions in each cell. Let $L(x)$ denote this approximately reconstructed function which can be written in this form
\begin{equation*}
  L(x)\ =\ \bar{\V}_i\ +\ \Sl_i\cdot\frac{x-x_i}{\Delta x}, \qquad
  x \in [x_{i-\frac12}, x_{i+\frac12}].
\end{equation*}
In order to $L(x)$ be a non-oscillatory approximation, we use the parabolic interpolation $\Q(x)$ constructed below to estimate the slopes $\Sl_i$ within each cell
\begin{equation*}
\Sl_i\ =\ \Delta x\times\minmod\Bigl(\od{\Q}{x}(x_i-0), \od{\Q}{x}(x_i+0)\Bigr).
\end{equation*}
In other words, the solution is reconstructed on the cells while the solution gradient is estimated on the dual mesh as it is often performed in more modern schemes \cite{Barth1994, Barth2004}. A brief summary of the UNO2 reconstruction can be also found in \cite{Dutykh2010, Dutykh2010e}.

% *********************************************************************************
\subsection{Treatment of the dispersive terms}
% *********************************************************************************

In this section, we explain how we treat numerically the dispersive terms of the Peregrine system \eqref{eq:per1}, \eqref{eq:per2} which are present only in the momentum conservation equation \eqref{eq:per2}. We propose the following approximation for the second component 
of $M(\bar{\V})$ of $\M(\bar{\V})$:
\begin{multline*}
  M_i(\bar{\V}) = \frac12 \bar{h}_i\frac{\bar{h}_{i+1}(\um_t)_{i+1} - 2\bar{h}_{i}(\um_t)_{i} + \bar{h}_{i-1}(\um_t)_{i-1}}{\Delta x^2} - \frac16 \bar{h}_i^2\frac{(\um_t)_{i+1} - 2(\um_t)_{i} + (\um_t)_{i-1}}{\Delta x^2} \\
  = \frac{\bar{h}_i}{2\Delta x^2}\Bigl(\bar{h}_{i-1} - \frac13 \bar{h}_i\Bigr)(\um_t)_{i-1} 
  - \frac{2}{3\Delta x^2}\bar{h}_i^2 (\um_t)_{i}
  + \frac{\bar{h}_i}{2\Delta x^2}\Bigl(\bar{h}_{i+1} - \frac13 \bar{h}_i\Bigr)(\um_t)_{i+1}.
\end{multline*}
Note that this spatial discretisation is of the second order $\O(\Delta x^2)$ so as to be consistent with the UNO2 advective flux discretisation presented above. If we denote by $I$ the identity matrix, we can now rewrite the semi-discrete scheme in the form
\begin{eqnarray*}
  \od{\hm}{t}\ +\ \frac{1}{\Delta x}\,\bigl[\,\F_{+}^{(1)}(\bar{\V}) 
  \,-\, \F_{-}^{(1)}(\bar{\V})\,\bigr]\ &=&\ 0, \\
  (I - M)\cdot\od{\um}{t}\ +\ \frac{1}{\Delta x}\,\bigl[\,
  \F_{+}^{(2)}(\bar{\V})
  \,-\, \F_{-}^{(2)}(\bar{\V})\,\bigr]\ &=&\ \S_b^{(2)},
\end{eqnarray*}
where $\F_{\pm}^{(1,2)}(\bar{\V})$ are the two components of the advective numerical flux vector $\F$ at the right ($+$) and left ($-$) faces correspondingly, and $\S_b^{(2)}$ denotes the discretisation of the second component of $\S_b$.

In order to advance the numerical solution forward in time one has to invert the matrix 
$(I - M)$ at every time step. This is no problem in practice, since the matrix appears to
be well conditioned in all cases we have considered. 
In fact, the invertibility of the matrix $(I - M)$ can be rigorously shown to hold 
for small enough $\Delta x$ since the matrix is then diagonally dominant. 
The criterion for diagonal dominance in the present case is seen
to be
$$
1 + \frac{2}{3\Delta x^2}\bar{h}_i^2 > 
  \left|-\frac{1}{6} \frac{\bar{h}_i^2}{\Delta x^2} + \frac{\bar{h}_i}{2 \Delta x^2}\bar{h}_{i-1}\right|
+ \left| -\frac{1}{6} \frac{\bar{h}_i^2}{\Delta x^2} + \frac{\bar{h}_i}{2 \Delta x^2}\bar{h}_{i+1} \right|.
$$
Using a Taylor expansion to express the terms $\bar{h}_{i-1}$ and $\bar{h}_{i+1}$ as
$\bar{h}_{i-1} = \bar{h}_i - \Delta x \bar{h}'(x_i) + \O(\Delta x^2)$ and
$\bar{h}_{i+1} = \bar{h}_i + \Delta x \bar{h}'(x_i) + \O(\Delta x^2)$ , respectively,
the criterion reduces to 
$$
1 + \frac{1}{3\Delta x^2}\bar{h}_i^2 > \frac{\bar{h}_i \bar{h}'_i}{\Delta x} +  \O(1),
$$
and this is guaranteed to hold for small enough $\Delta x$.

% *********************************************************************************
\subsection{Time stepping}
% *********************************************************************************

We assume that the linear system of equations is already inverted and we have the following system of ODEs:
\begin{equation*}
  \vm_t = \N (\vm, t), \qquad \vm(0) = \vm_0.
\end{equation*}
In order to solve numerically the last system of equations, we apply the Bogacki-Shampine method proposed by Przemyslaw Bogacki and Lawrence F.~Shampine in 1989 \cite{Bogacki1989}. It is a Runge-Kutta scheme of the third order with four stages. It has an embedded second order method which is used to estimate the local error and thus, to adapt the time step size. Moreover, the Bogacki-Shampine method enjoys the First Same As Last (FSAL) property so that it needs approximately three function evaluations per step. This method is also implemented in the \texttt{ode23} function in Matlab \cite{Shampine1997}. The one step of the Bogacki-Shampine method is given by:
\begin{eqnarray*}
  k_1 &=& \N (\vm^{(n)}, t_n), \\
  k_2 &=& \N (\vm^{(n)}+\half \Delta t_n k_1, t_n + \half\Delta t), \\
  k_3 &=& \N (\vm^{(n)})+\frth \Delta t_n k_2, t_n + \frth\Delta t), \\
  \vm^{(n+1)} &=& \vm^{(n)} + \Delta t_n\bigl(\textstyle{2\over9}k_1 + \textstyle{1\over3}k_2 + \textstyle{4\over9}k_3\bigr), \\
  k_4 &=& \N(\vm^{(n+1)}, t_n + \Delta t_n), \\
  \vm_2^{(n+1)} &=& \vm^{(n)} + \Delta t_n\bigl(\textstyle{4\over{24}}k_1 +
  \textstyle{1\over4}k_2 + \textstyle{1\over3} k_3 + \textstyle{1\over8}k_4\bigr).
\end{eqnarray*}
Here $\vm^{(n)} \approx \vm(t_n)$, $\Delta t$ is the time step and $\vm_2^{(n+1)}$ is a second order approximation to the solution $\vm(t_{n+1})$, so the difference between $\vm^{(n+1)}$ and $\vm_2^{(n+1)}$ gives an estimation of the local error. The FSAL property consists in the fact that $k_4$ is equal to $k_1$ in the next time step, thus saving one function evaluation.

If the new time step $\Delta t_{n+1}$ is given by $\Delta t_{n+1} = \rho_n\Delta t_n$, 
then according to the H211b digital filter approach \cite{Soderlind2003, Soderlind2006}, the proportionality factor $\rho_n$ is given by
\begin{equation}\label{eq:tadapt}
  \rho_n = \Bigl(\frac{\delta}{\epsilon_n}\Bigr)^{\beta_1} \Bigl(\frac{\delta}{\epsilon_{n-1}}\Bigr)^{\beta_2}\rho_{n-1}^{-\alpha},
\end{equation}
where $\epsilon_n$ is a local error estimation at time step $t_n$, and the constants $\beta_1$, $\beta_2$ and $\alpha$ are defined by
\begin{equation*}
  \alpha = \frac14, \quad \beta_1 = \frac{1}{4p}, \quad \beta_2 = \frac{1}{4p}.
\end{equation*}
The parameter $p$ gives the order of the scheme, and $p = 3$ in our case.

\begin{remark}
The adaptive strategy \eqref{eq:tadapt} can be further improved if we smooth the factor $\rho_n$ before computing the next time step $\Delta t_{n+1}$:
\begin{equation*}
  \Delta t_{n+1} = \hat\rho_n\Delta t_n, \qquad
  \hat\rho_n = \omega(\rho_n).
\end{equation*}
The function $\omega(\rho)$ is called \emph{the time step limiter} and should be smooth, monotonically increasing and should satisfy the following conditions:
\begin{equation*}
  \omega(0) < 1, \quad \omega(+\infty) > 1, \quad \omega(1) = 1, \omega'(1) = 1.
\end{equation*}
One possible choice was suggested in \cite{Soderlind2006}:
\begin{equation*}
  \omega(\rho) = 1 + \kappa\arctan\Bigl(\frac{\rho-1}{\kappa}\Bigr).
\end{equation*}
In our computations the parameter $\kappa$ is set to 1.
\end{remark}

\begin{figure}
\centering
\subfigure[]{\includegraphics[width=0.49\textwidth]{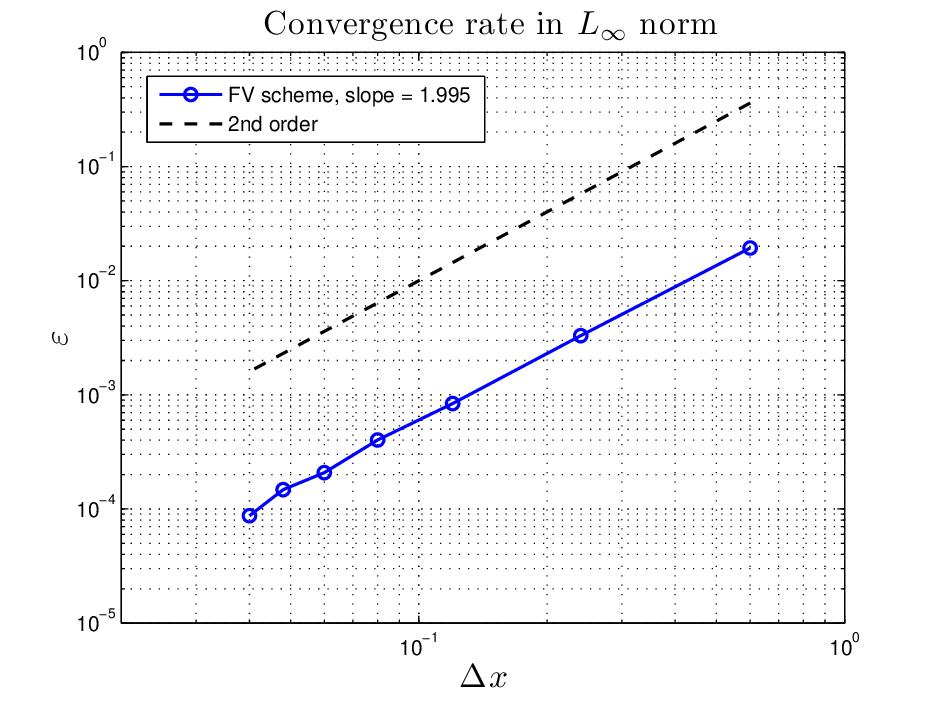}}
\subfigure[]{\includegraphics[width=0.49\textwidth]{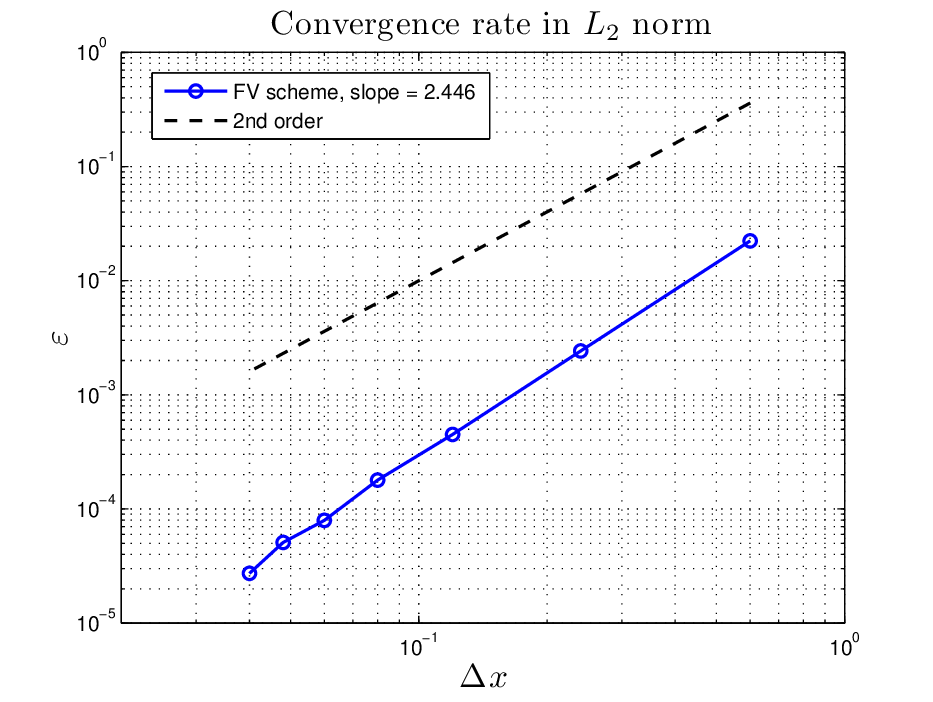}}
\caption{\small\em Convergence rate of the finite-volume scheme in the $L^{\infty}$-norm (left panel) and the  $L^2$-norm (right panel). The numerical integration of a solitary wave as shown in Figure \ref{Grimshaw} is compared to a translated profile. It appears that the second-order convergence is achieved.}
\label{Convergence}
\end{figure}

% *********************************************************************************
\subsection{Validation}
% *********************************************************************************

The scheme described in this section is implemented in MATLAB, and runs on a workstation. To check whether the implementation is correct, we use the approximate solitary waves of \eqref{Peregrine}, computed in the last section. These are used as initial data in the fully discrete scheme, and integrated forward in time. The computed solutions are then compared to the same solitary waves, but shifted forward in space by $ct_0$, where $c$ is the wave speed, and $t_0$ is the final time. This procedure is repeated a number of times with different spatial gridsizes. As a result, it is possible to find the spatial convergence rate of the scheme. As is visible in Figure~\ref{Convergence}, the convergence achieved by the practical implementation of the discretisation is very close to the theoretical convergence rate. Since the temporal discretisation is adaptive, we do not present a convergence study in terms of the timestep $\Delta t$.

% *********************************************************************************
\subsubsection{Wave generation by moving bottom}
% *********************************************************************************

We have just shown the convergence of our scheme under the mesh refinement. Even if the solution we used in validation is fully nonlinear, it only exists on the flat bottom. Since in the present study we are mainly interested in the wave generation by bottom motion, the next validation test will be entirely devoted to this question. Namely, we are going to use an analytical solution to the linearized full Euler equations also known as the Cauchy--Poisson problem. The use of this solution in tsunami generation problems was first proposed by J.~\textsc{Hammack} (1973) \cite{Hammack}.

\begin{figure}
  \centering
  \subfigure[$t = 1.5$]%
  {\includegraphics[width=0.59\textwidth]{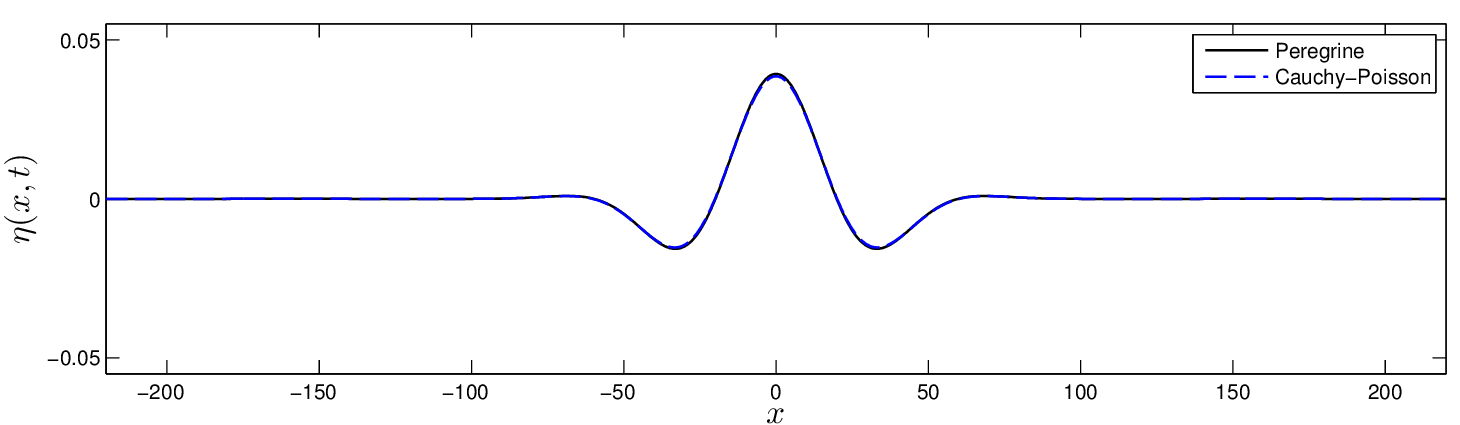}}
  \subfigure[$t = 4.0$]%
  {\includegraphics[width=0.59\textwidth]{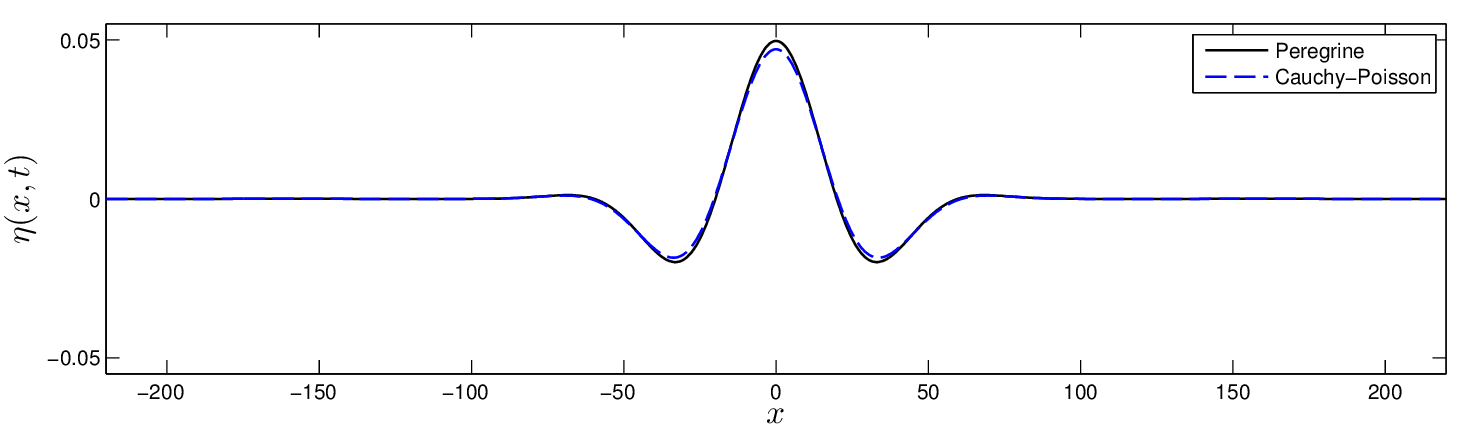}}
  \subfigure[$t = 8.0$]%
  {\includegraphics[width=0.59\textwidth]{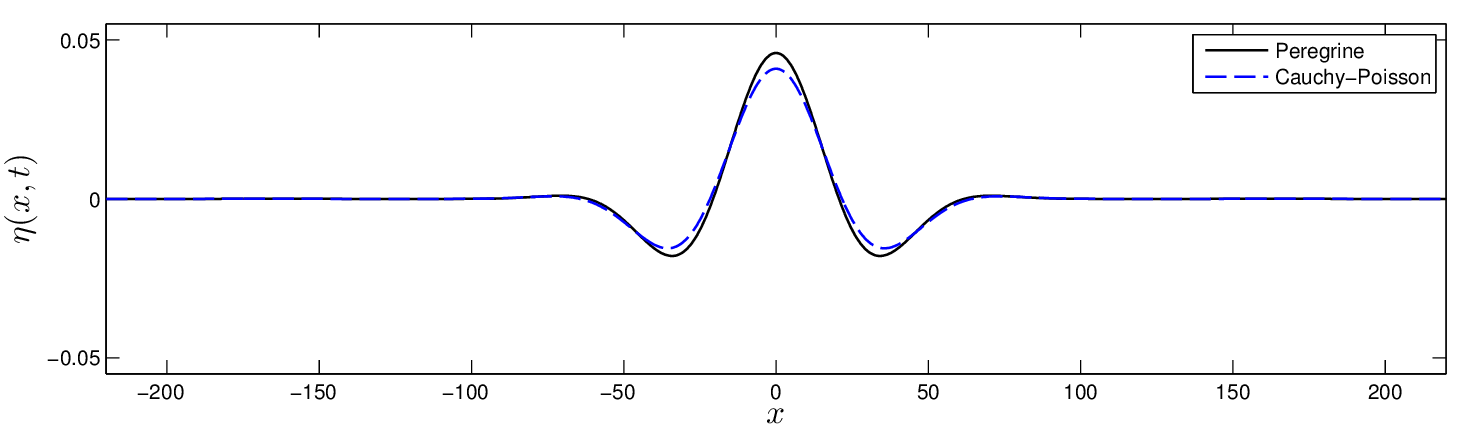}}
  \subfigure[$t = 20.0$]%
  {\includegraphics[width=0.59\textwidth]{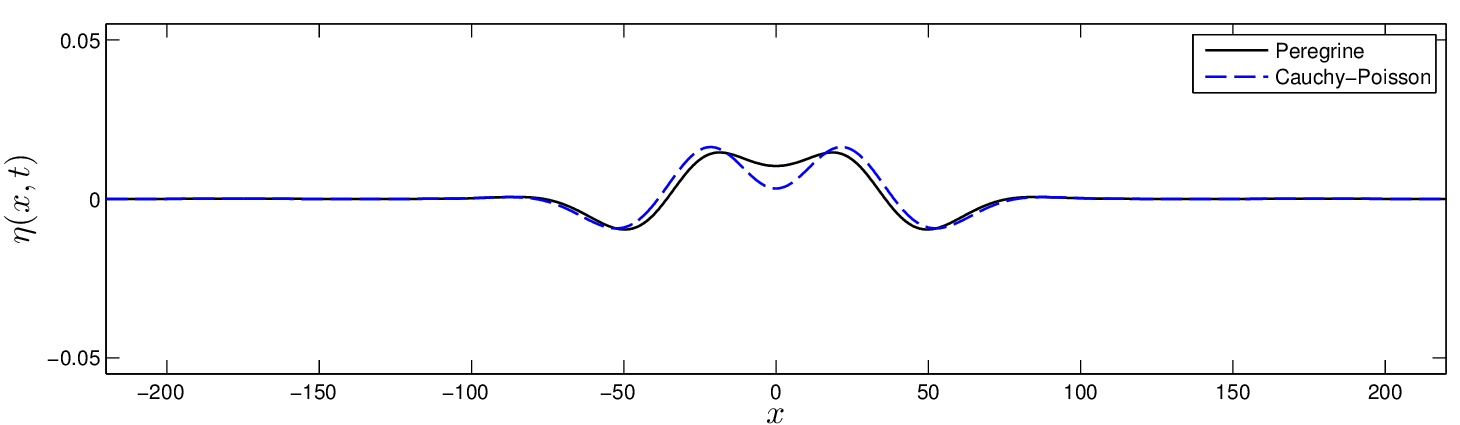}}
  \subfigure[$t = 80.0$]%
  {\includegraphics[width=0.59\textwidth]{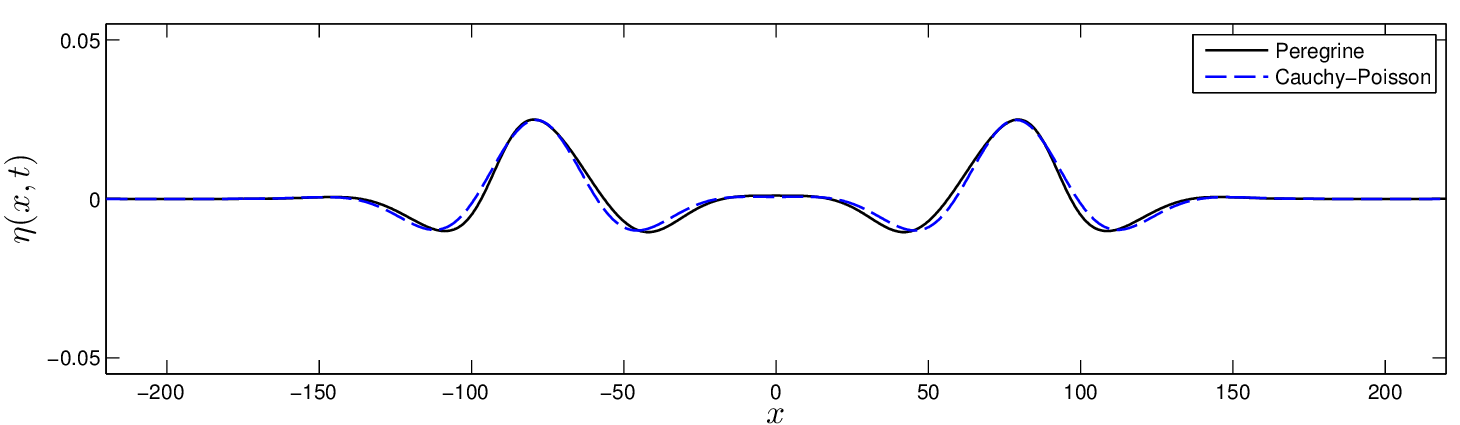}}
  \caption{\small\em Free surface waves generated by a moderately fast bottom motion. The blue dashed line corresponds to the analytical Cauchy--Poisson solution, while the solid black line is our numerical solution to the Peregrine system.}
  \label{fig:cp_slow}
\end{figure}

\begin{figure}
  \centering
  \subfigure[$t = 2.5$]%
  {\includegraphics[width=0.59\textwidth]{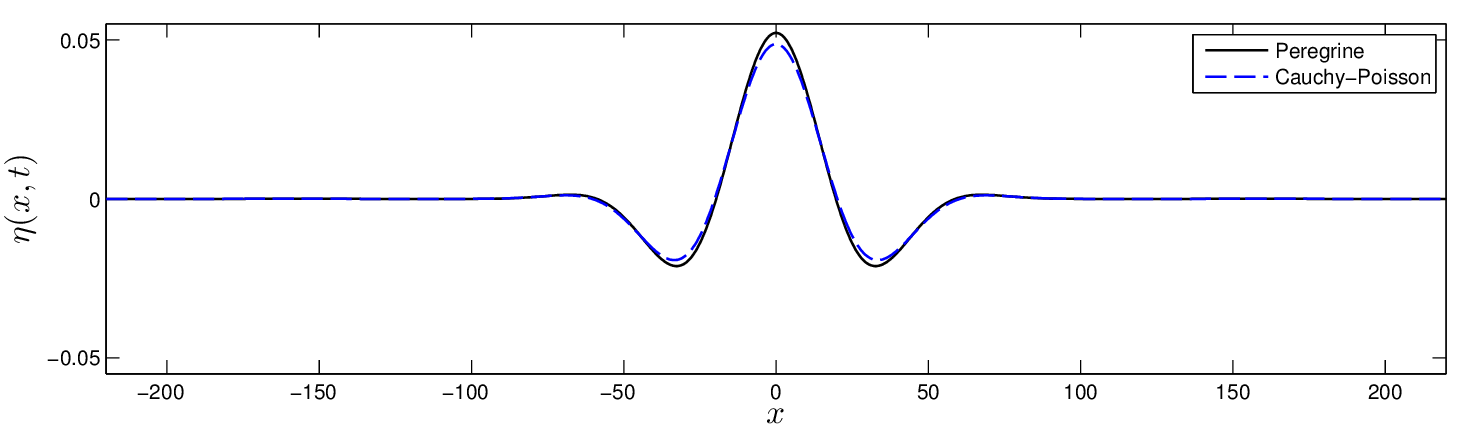}}
  \subfigure[$t = 8.0$]%
  {\includegraphics[width=0.59\textwidth]{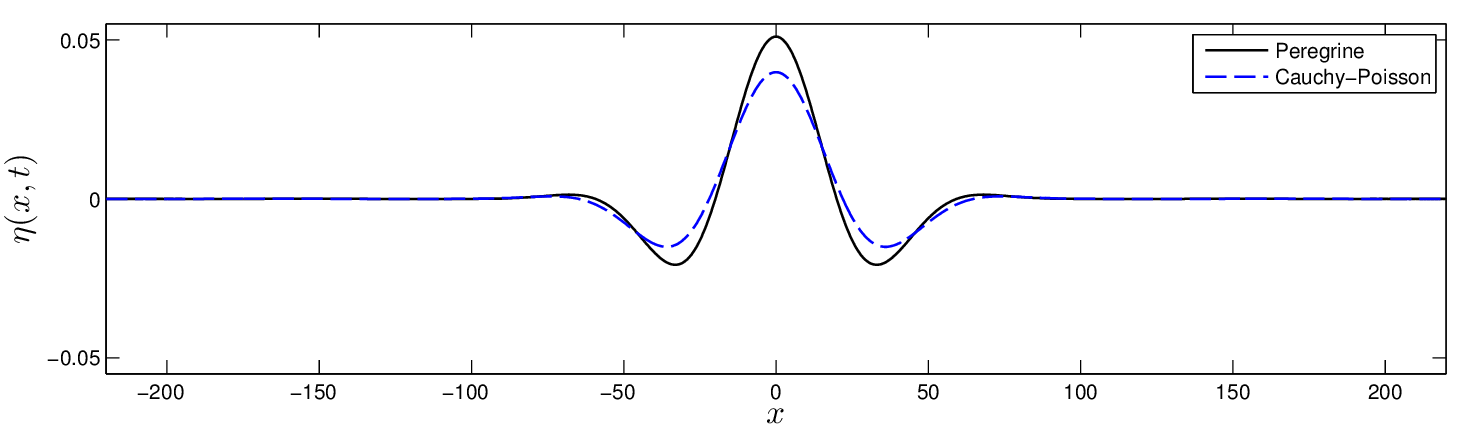}}
  \subfigure[$t = 15.0$]%
  {\includegraphics[width=0.59\textwidth]{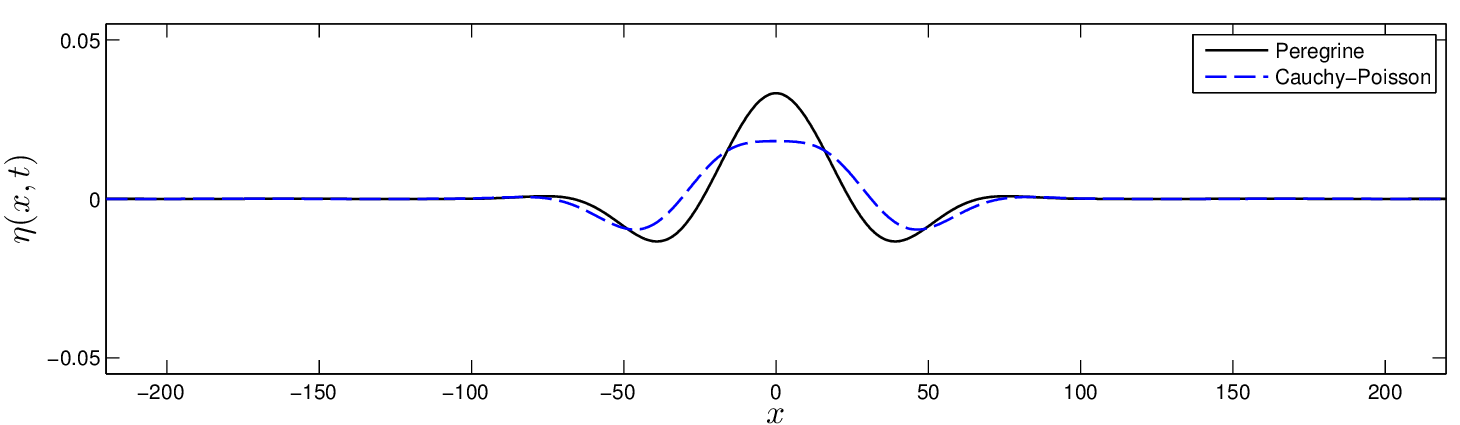}}
  \subfigure[$t = 25.0$]%
  {\includegraphics[width=0.59\textwidth]{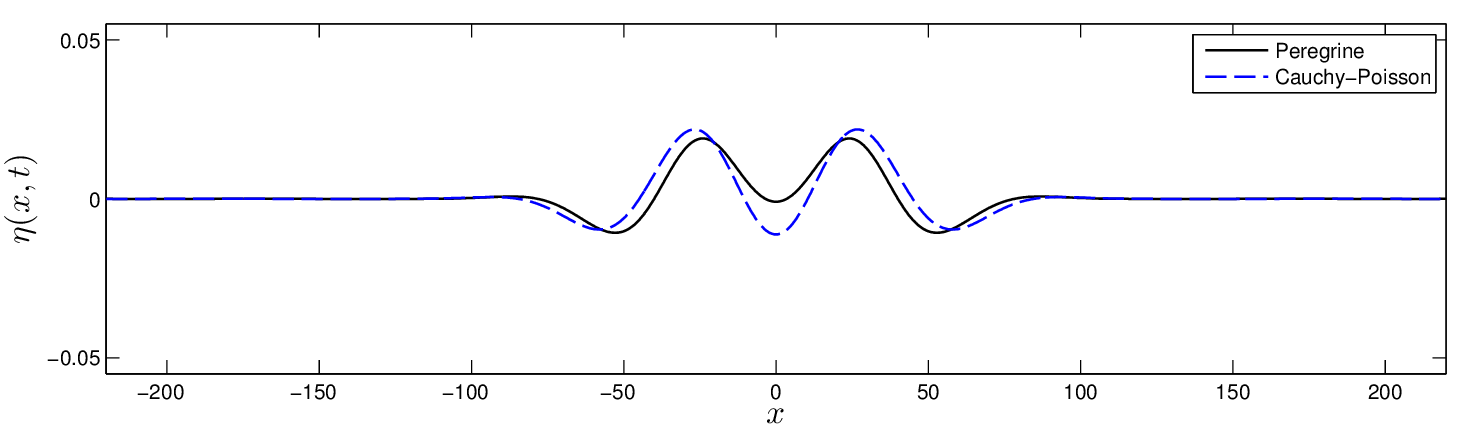}}
  \subfigure[$t = 70.0$]%
  {\includegraphics[width=0.59\textwidth]{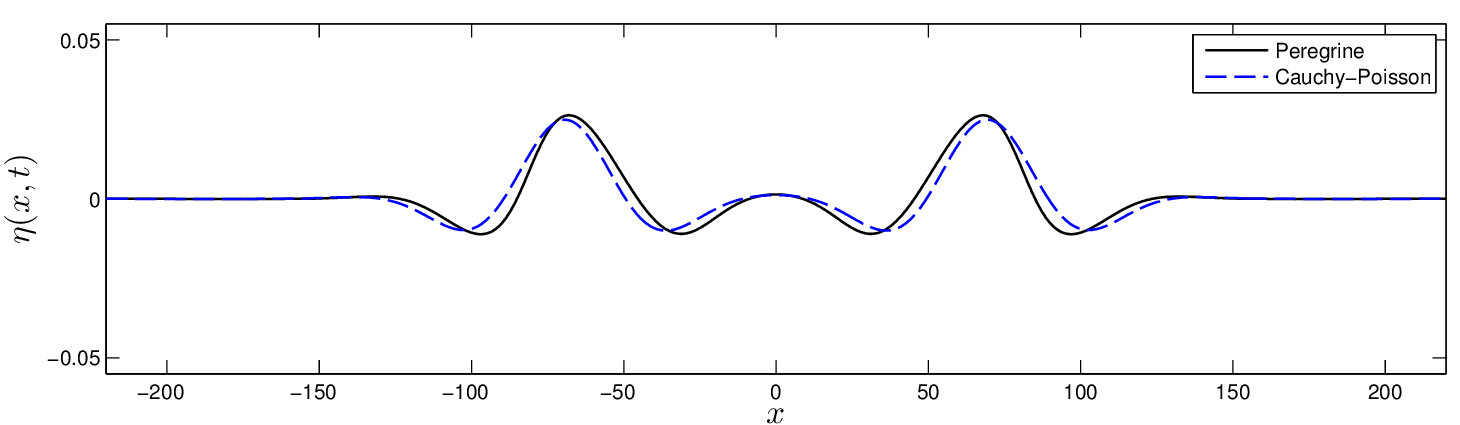}}
  \caption{\small\em Free surface waves generated by a fast bottom motion. The blue dashed line corresponds to the analytical Cauchy--Poisson solution, while the solid black line is our numerical solution to the Peregrine system.}
  \label{fig:cp_fast}
\end{figure}

We consider the linearized water wave problem for a fluid layer of uniform depth $z = -d_0 = \mathrm{const}$. However, a portion of the bottom can move vertically and its deformation is given by a smooth function $\zeta(x,t)$ such that $\zeta(x,0) \equiv 0$ at time $t$ the bottom profile is given by $z = -d_0 + \zeta(x,t)$. Moreover, we will make a special assumption about the structure of the bottom deformation:
\begin{equation*}
  \zeta(x,t) = T(t)\zeta_0(x), \qquad T(t) := 1 - \ue^{-\alpha t}, \qquad \alpha > 0, \quad t \geq 0.
\end{equation*}
Obviously, we have to assume that $||\zeta_0|| \ll 1$ so that the linear approximation be valid. Then, the free surface elevation at any time is given by the following formula \cite{Hammack,Dutykh2009c}:
\begin{equation}\label{eq:CP}
  \eta(x,t) = -\frac{\alpha^2}{2\pi}\int_{\R} \frac{\hat{\zeta}_0(k)}{\cosh(kd_0)}\cdot \frac{\ue^{-\alpha t} - \cos(\omega t) - \frac{\omega}{\alpha}\sin(\omega t)}{\alpha^2 + \omega^2} \cdot \ue^{-\ui k x}\,\ud k,
\end{equation}
where $\hat{\zeta}_0(k)$ is the Fourier transform of $\zeta_0(x)$ and $\omega := \sqrt{gk\tanh(kd_0)}$ is the wave frequency corresponding to the wavenumber $k$. The above integral can be easily computed using the FFT algorithm. To fix the ideas for numerical computations, we will take the following localized oscillatory bottom deformation:
\begin{equation*}
  \zeta_0(x) = a\cos(k_0x)\ue^{-\lambda_0 x^2}, \qquad \lambda_0 > 0.
\end{equation*}
The values of all parameters used in numerical simulation are given in Table~\ref{tab:bot}. The nonlinearity parameter $a/d_0$ is chosen to be $0.05$, which is far above the nonlinearity of the earthquake generated tsunamis. However, we think that this value corresponds better to the scope of the present study. In order to simulate this set-up using the Peregrine system, we consider a symmetric 1D computational domain $[-220, 220]$ discretized into $N = 2000$ equal control volumes. The time stepping tolerance parameter was set far below the spatial discretization error ($\sim \O(\Delta x^2)$). First, we will take a moderately fast bottom uplift corresponding to the parameter $\alpha = 1.0$. Computational results are presented in Figures~\ref{fig:cp_slow}(a--e). One can see that the overall agreement is fairly good even if some small differences can be noticed on Figures~\ref{fig:cp_slow}(c-d). However, the resulting wave form predicted by the Peregrine system follows closely the linearized full Euler solution \eqref{eq:CP}. Now, we will double the bottom uplift speed ($\alpha = 2.0$). This result is presented in Figures~\ref{fig:cp_fast}(a--e). One can see more substantial differences during the generation phase (see panels b--d). However, here again the resulting wave is surprisingly well represented by the Boussinesq-type equations. The observed discrepancies during the generation phase are essentially due to the simplified structure of the vertical speed in Boussinesq-type equations \cite{Dutykh2010a}.
\begin{table}
  \centering
  \begin{tabular}{l|c}
  \hline\hline
  Gravity acceleration: $g$ & 1.0 \\
  Undisturbed water depth: $d_0$ & 1.0 \\
  Bottom displacement amplitude: $a$ & 0.05 \\
  Bottom oscillation inverse length: $k_0$ & $\frac{\pi}{40}$ \\
  Bottom localization parameter: $\lambda_0$ & $0.7\times 10^{-3}$ \\
  Vertical uplift speed: $\alpha$ & $1.0$ and $2.0$ \\
  \hline\hline
  \end{tabular}
  \bigskip
  \caption{\small\em Values of various parameters used to simulate the wave generation by moving bottom.}
  \label{tab:bot}
\end{table}
%
% *********************************************************************************
\section{Numerical results and discussion}\label{Results}
\begin{figure}
\centering
\subfigure[]{\includegraphics[width=0.49\textwidth]{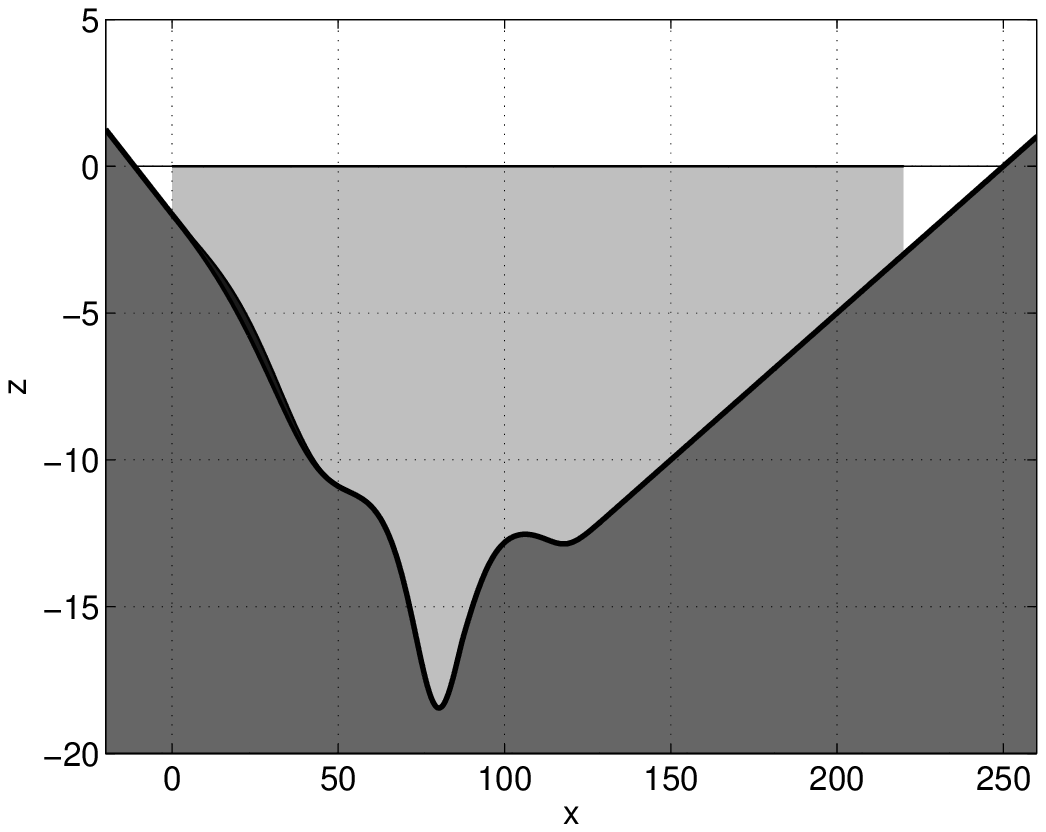}}
\subfigure[]{\includegraphics[width=0.49\textwidth]{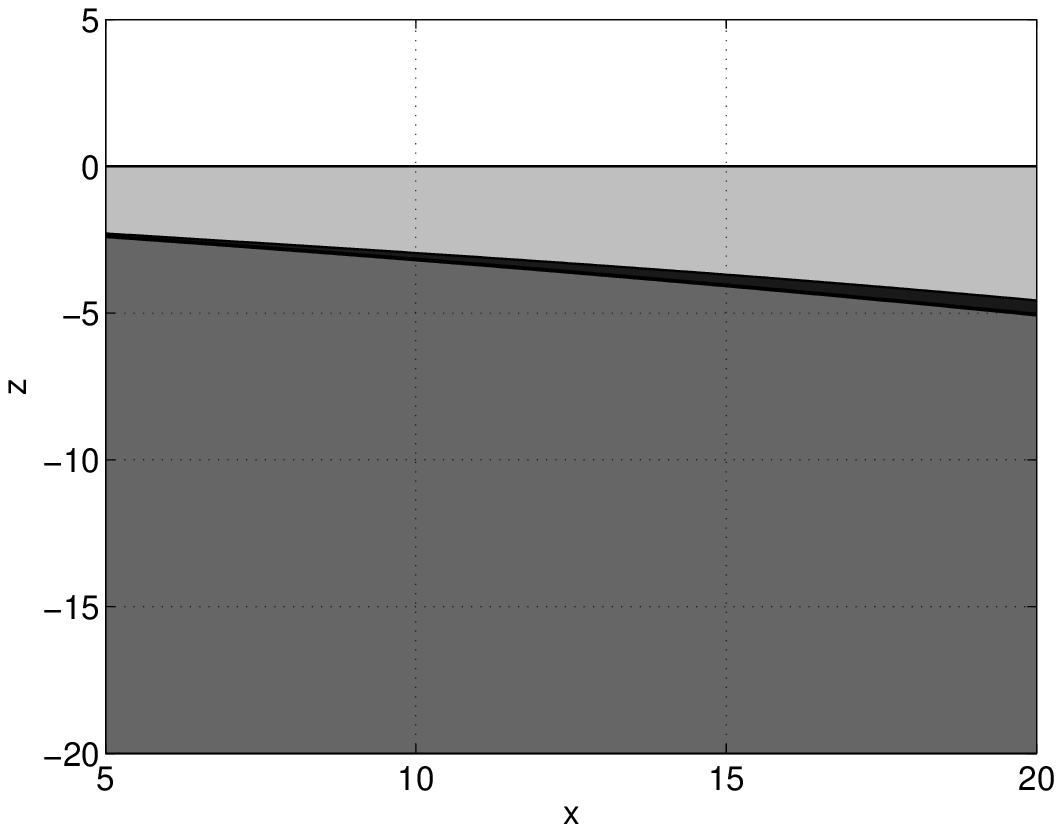}}
\caption{\small\em The physical setup of the problem. The river bed is indicated in dark grey. The computational fluid domain is shaded light grey, and the landslide is visible in black. Note the difference in horizontal and vertical scales in the left panel. The right panel shows a closeup of the left beach and part of the landslide in a one-to-one aspect ratio.}
\label{Setup}
\end{figure}

% *********************************************************************************
%
Let us consider a one-dimensional computational domain $I = [a, b] = [0, 220]$ composed of two regions: the generation region and a sloping beach on the right. More specifically, the static bathymetry function $h_0(x)$ is given by a smoothed out profile generated from the expression
\begin{equation*}
  h_0(x) = \left\{
  \begin{array}{ll}
    d_0 + \tan\delta\cdot (x-a) + p(x), & a \leq x \leq m, \\
    d_0 + \tan\delta\cdot (m-a) - \tan\delta\cdot (x-m), & x > m, \\
  \end{array}
  \right.
\end{equation*}
where the function $p(x)$ is defined as
\begin{equation*}
  p(x) =  A_1\sech(k_1(x-x_1)) + A_2\sech(k_2(x-x_2)),
\end{equation*}
In essence, this function represents a perturbation of the sloping bottom by two underwater bumps. We made this nontrivial choice in order to illustrate the advantages of our landslide model, which was designed to handle general non-flat bathymetries. The parameters can be chosen in order to fit a given bathymetry, but the particular values used here are $A_1 = 4.75$, $A_2 = 8.85$, $k_1 = 0.06$, $k_2 = 0.13$, $x_1 = 45$, $x_2 = 80$, and $m = 120$. The bottom profile for these parameters is depicted in Figure~\ref{Setup}. Of course, in general, if the bottom topography is known, then a numerical bathymetry map could also be used.

\begin{table}
\centering
\begin{tabular}{c|l|c|cc}
Symbol     &   Parameter & Units & Values &  \\
\hline\hline
$g$   &  gravitational acceleration &  $\mbox{m/s}^2$   & $9.81$  & \\
$d_0$ &  water depth at $x=a$       &   $\mbox{m}$ & $1.0-2.0$    & \\
$\tan(\delta)$ &  bottom slope     &  & $0.1$              & \\
$A$  &   landslide amplitude  &     $\mbox{m}$  & $0.55$          & \\
$l$  & landslide length    &  $\mbox{m}$ &  $52.4$ & \\
$c_w$  &   added mass coefficient     &        & $1.0$ & \\
$c_d$  &   water drag coefficient    &    & $1.0$ &  \\
$c_f$  &   friction coefficient      &     & $\tan(3^{\circ})$ &  \\
$\gamma$ & density ratio water/landslide  &  & $1.8$  &\\
$c_b$  &   friction coefficient with bottom &   & $7.63\times 10^{-4}$ &  \\
$c_v$  &   viscous friction coefficient    &     & $1.27\times 10^{-3}$ & \\
   \hline\hline
\end{tabular}
\vskip 0.1in
\caption{\small\em Values of various parameters used in numerical computations.}
\label{tab:pars}
\end{table}

\begin{figure}
\centering
\includegraphics[width=0.69\textwidth]{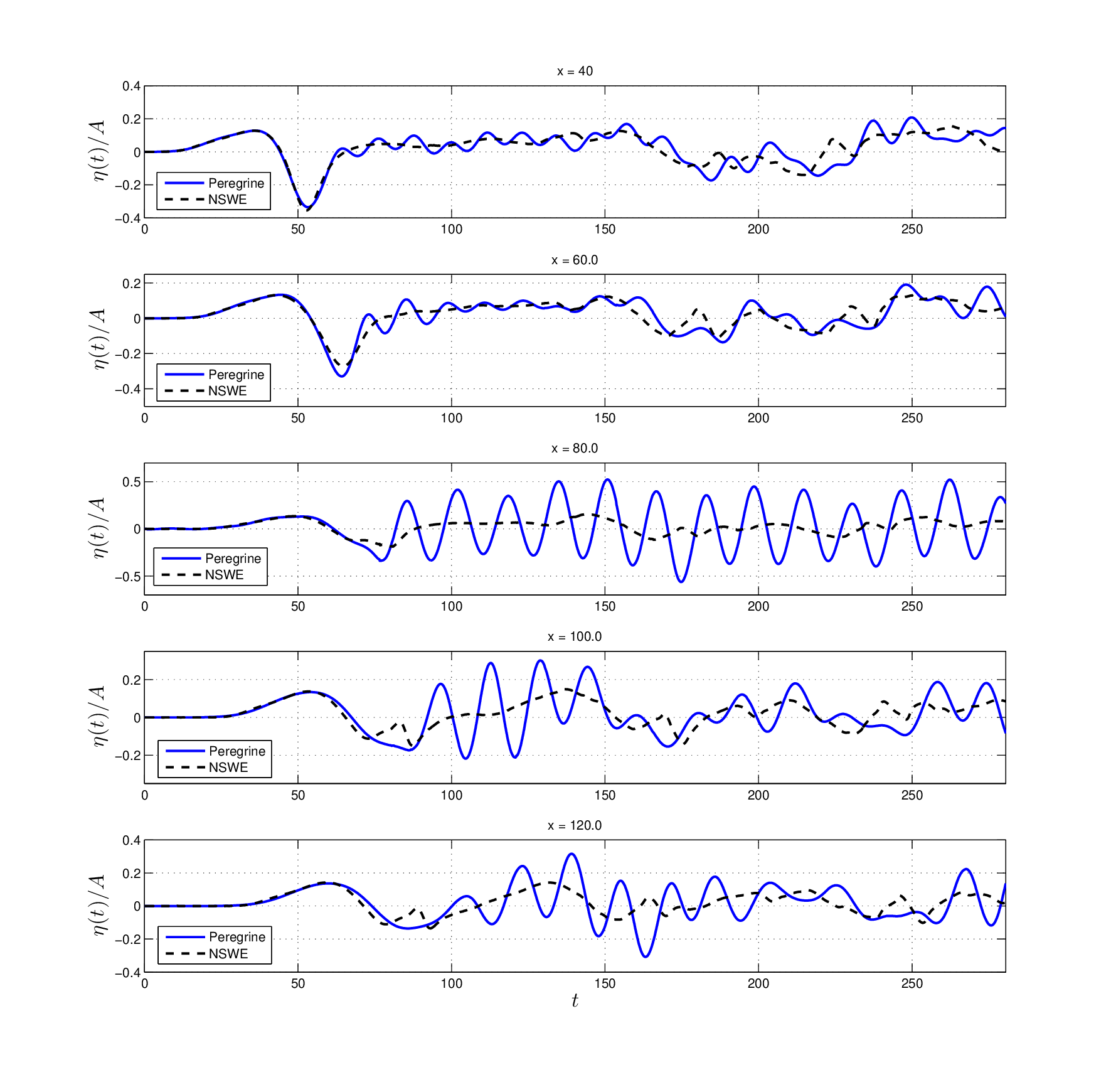}
\caption{\small\em Time series of the surface elevation at wave gauges located at $x=40$, $x=60$, $x=80$, $x=100$ and $x=120$. The solid (blue) curve depicts the wave elevation computed with the dispersive system \eqref{PeWu}, and the dashed curve represents results obtained from the shallow-water system. All variables are non-dimensional.}
\label{GaugesA}
\end{figure}

We now present some results of the solution of the surface wave problem using the model in Section~\ref{BoussinesqModel}, integrated numerically with the method of Section~\ref{NumericalScheme}. A landslide is introduced on the left side of the bathymetry, and using the method of Section~\ref{LandslideModel}, its path along the bottom is determined by following the barycenter. Simultaneously, the system \eqref{PeWu} is solved with the time-dependent bottom topography given from the solution of the landslide problem. The problem is integrated up to a final time $T$. Figure~\ref{GaugesA} shows wave records at six virtual wave gauges for both the dispersive system \eqref{PeWu} and the shallow-water system. It appears from this figure that the shallow-water system underpredicts the development of free-surface oscillations. In particular, the wave gauges located at $x = 40$ and $x = 60$ show similar waveheights for both the shallow-water, and the dispersive system, but a qualitative divergence, as small oscillations are already developing which are not captured by the shallow-water system. Once the waves have propagated to the wave gauges located at $x = 80$, the dispersive oscillations have amplified, so that the waveheight is larger by a factor of $2$ to $3$ than the waveheight predicted by the shallow-water system. Going further to the wave gauges located at $x = 100$ and $x = 120$, the now rising bottom starts to have a damping effect on the waves.

\begin{figure}
\centering
\includegraphics[width=0.69\textwidth]{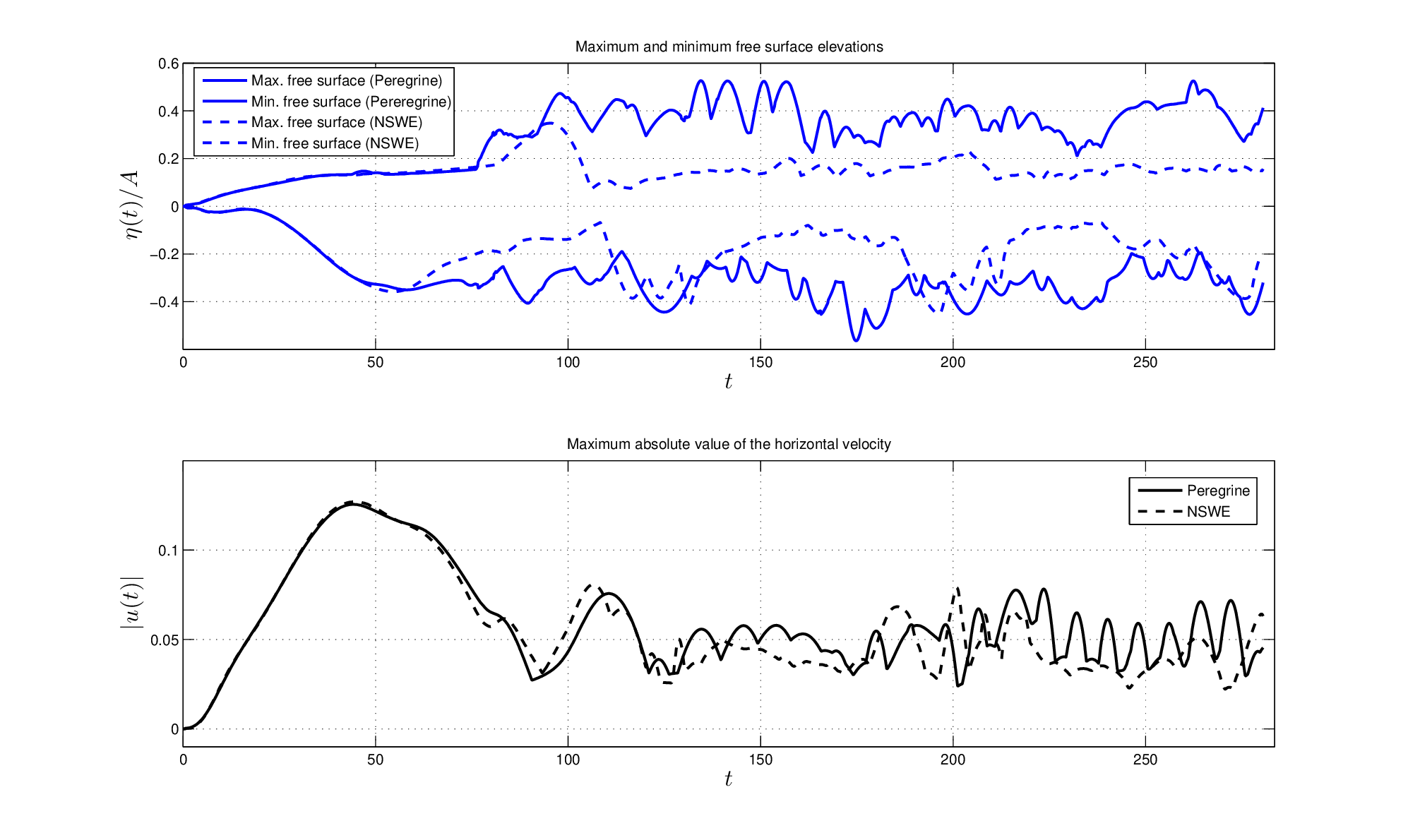}
\caption{\small\em Maximum and minimum of the surface excursion, and the horizontal velocity as a function of (non-dimensional) time.}
\label{MaxMinVel}
\end{figure}

The maximum and minimum free surface elevation over the whole domain is shown in Figure~\ref{MaxMinVel}. On the lower panel of the same Figure~\ref{MaxMinVel} we show the maximal unsigned horizontal velocity. One can see that for short times the hydrostatic and dispersive models give very close extreme values. Later the differences start to appear due to the accumulation of dispersive effects.

\begin{figure}
\centering
\includegraphics[width=0.69\textwidth]{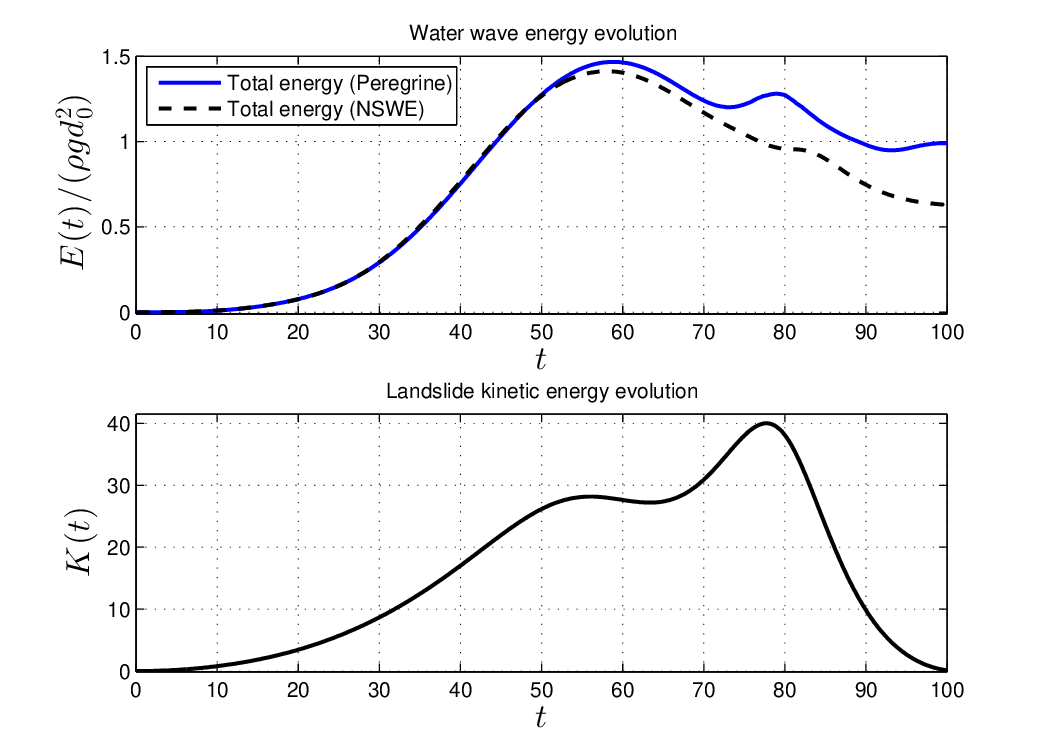}
\caption{\small\em Development of the wave energy, and the kinetic energy of the landslide as a function of (non-dimensional) time. Note that the kinetic energy of the landslide starts from $0$ (all energy is potential) and also ends at $0$ (all energy has been dissipated or transferred to the fluid).}
\label{Energy}
\end{figure}

Figure~\ref{Energy} shows the development of the kinetic energy of the landslide mass and simultaneously the total (kinetic plus potential) energy contained in the body of the fluid and the surface waves. Energy development is an important question in the study of tsunamis, and there have been studies exclusively devoted to this question \cite{Tinti2000}. Energy issues connected to water wave models of Boussinesq type have also been studied before \cite{Ali2010, AliKalisch2012, Dutykh2009b}. While these models contained a source of energy, in the case at hand, the work done by friction as the landslide slides down the bottom acts as a drain of energy, and after the landslide has come to rest, all energy has been transferred to the fluid. However, not all energy can be considered as residing in the wave motion, because a significant amount of energy is needed to lift the water from the final position of the landslide to the initial position of the landslide. This results in a large increase in potential energy of the fluid, and only a fraction of the potential energy of the landslide is transferred to the wave motion. This fact has also been explained in previous works \cite{Harbitz2006}.

In order to compute the wave energy in the fluid, we use the integral
\begin{equation}
E_{w} = \int_a^b  \left\{ \frac{g}{2} \eta^2  + \half(h_0 + \eta) u^2 \right\} \,\ud x,
\end{equation}
which arises from the shallow-water theory. The kinetic energy of the landslide is given by
\begin{equation}
E_{sl} = \half m v^2,
\end{equation}
with the generalised mass $m$ given by \eqref{genMass}, and $v = \od{s}{t}$ as defined in Section~\ref{LandslideModel}. Figure~\ref{Energy} shows the development of the wave energy and kinetic energy of the landslide. The upper panel shows the energy according to the shallow-water and dispersive model. The lower panel shows the kinetic energy of the landslide.

We have also computed the Froude number $\Fr = \frac{v}{\sqrt{gh(x_c)}}$ during the evolution. Here $v$ is the $x$-component of the velocity of the barycenter of the landslide, $x_c$ is the position of the barycenter, and $h(x_c)$ is the corresponding local water depth. This number was always found to be much less than $1$ in all numerical experiments. The maximum value was generally about $0.5$.

\begin{figure}
\centering
\includegraphics[width=0.69\textwidth]{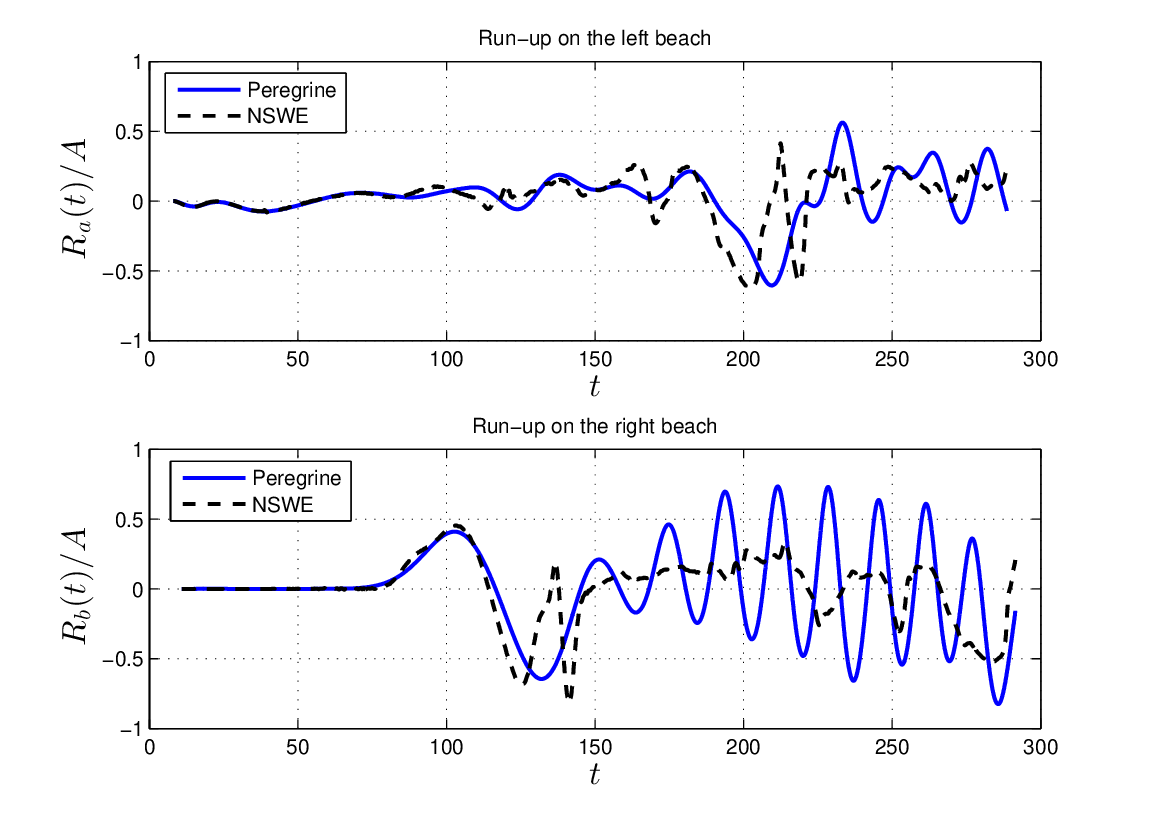}
\caption{\small\em Run-up on the left and right beach using \eqref{Choi}, computed with the dispersive system (solid curve) and the nonlinear shallow-water system (dashed curve) as a function of (non-dimensional) time.}
\label{runup}
\end{figure}

To compute the wave run-up and draw-down, we use exact representations given by Choi \emph{et al}. \cite{Choi2011} (a similar formula was also derived in \cite{Didenkulova2008}). On the right beach, the undisturbed water depth at the edge of the computational domain is $h=3$, and the distance from the computational domain to the shore line is $L=30$. Using the shallow-water wave speed, the travel time of a wave from the edge of the computational domain to the shore is computed as 
\begin{equation}\label{T}
  T = \frac{2L}{\sqrt{gh}} = 2 \sqrt{\frac{L}{g \alpha}}.
\end{equation}
Then the formula for the wave run-up $R$ at the shore reads
\begin{equation}\label{Choi}
R = \int_0^{t-T} \frac{t-\tau}{(t-\tau)^2 - T^2}\cdot\od{\eta}{\tau}(x,\tau) \, \ud\tau
\end{equation}
with $x=220$. At the left beach, the undisturbed water depth is $h = 1.642$, and the distance to the beach is $L=11.2814$. A similar formula can be then be computed for $x=0$.

Figure~\ref{runup} shows the run-up on the left and right beaches both in the Boussinesq scaling and in the shallow-water theory. While the agreement is fair on the left beach, it appears immediately that the Boussinesq theory predicts a wave run-up on the right beach which is much larger (roughly by a factor of two) than the wave run-up according to the shallow-water theory. A possible explanation for this divergence is the nature of the numerical solver when applied to the shallow-water system. In this case, there is continuous numerical dissipation through the handling of hyperbolic wave breaking. Since the waves do not break in the Boussinesq scaling, the dissipation is not present, or at least much smaller. The difference can also be read off from the comparison of the wave energy in the Boussinesq and shallow-water system provided in Figure~\ref{Energy}. It can be seen there that the wave energy in the shallow-water model starts to diverge from the Boussinesq model at non-dimensional time $t = 50$. The difference between the two increases continuously, until at the final time, the Boussinesq energy is about $50 \%$ larger than the shallow-water energy. Note that significant run-up in Figure~\ref{runup} does not happen until non-dimensional time $t=75$, at which time the energy in the Boussinesq system is already much larger than in the shallow-water system.

\begin{figure}
\centering
\includegraphics[width=0.69\textwidth]{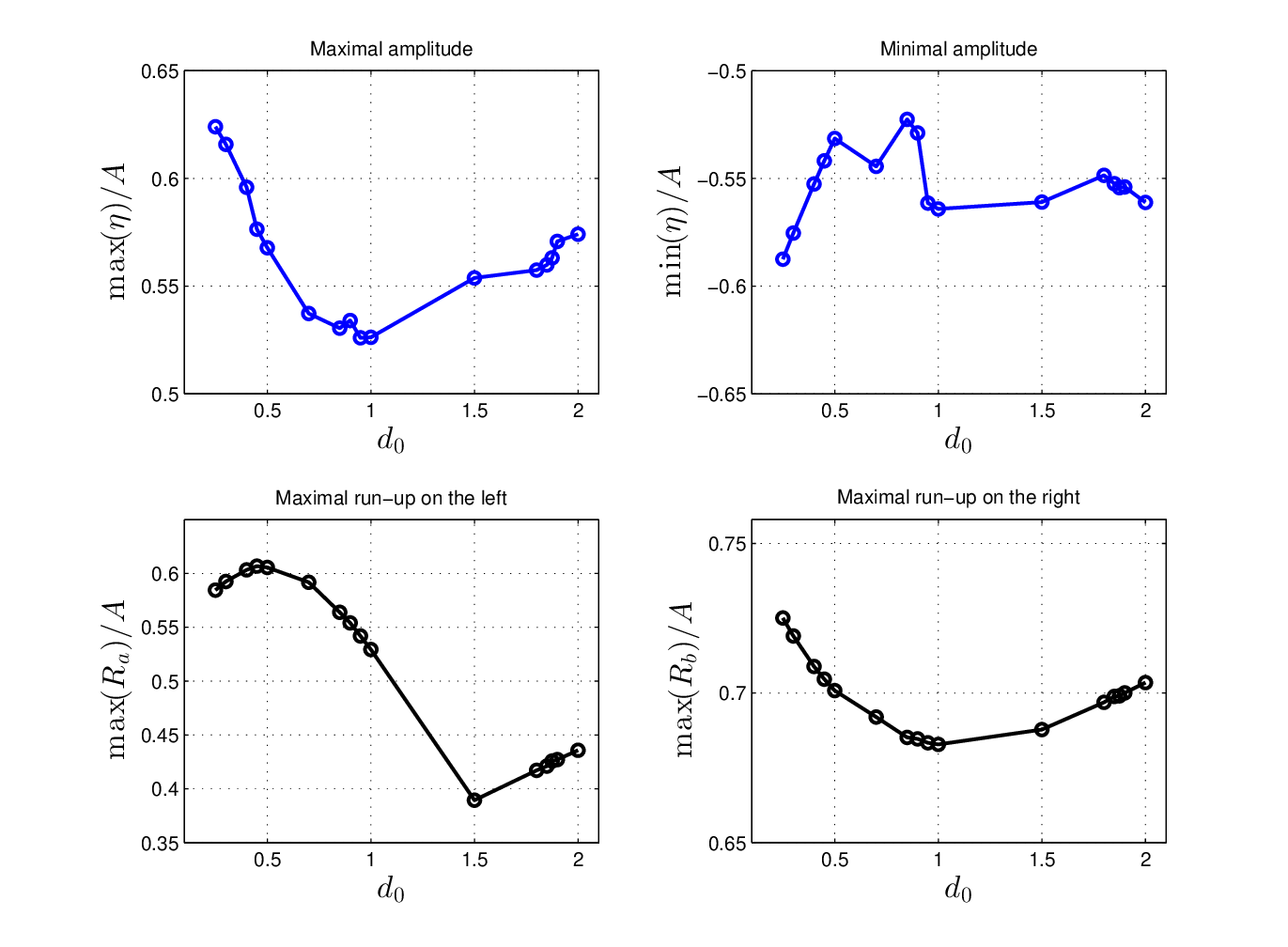}
\caption{\small\em Maximal and minimal wave amplitude, and the maximum run-up on the left and right beaches as a function of the initial depth of the center of the landslide $d_0$.}
\label{MaxMinAmp}
\end{figure}

In Figure~\ref{MaxMinAmp}, we have plotted the maximum wave amplitude, the minimum wave amplitude, and the maximum wave run-up on the left and right beaches. In comparison to previous studies, such as \cite{Grilli2005}, where an open domain was used, it appears that in our case, the maximal amplitude, as well as the run-up have a minimum at $d_0$ between $1$ and $1.5$. In \cite{Grilli2005}, it was found that maximum wave amplitude and run-up (on the left beach) were strictly decreasing functions of $d_0$. The phenomenon of rising amplitude and run-up may be accredited to resonant effects which are absent on an open domain (such as an ocean beach), but cannot be neglected for tsunamis generated by landslides in rivers and lakes.

% **********************************************************************************
\section{Conclusion}\label{Conclusion}
% **********************************************************************************

The influence of an underwater landslide on surface waves in a closed basin has been studied. The key features of the study have been that the motion of the underwater landslide is determined by integrating a second-order ordinary differential equation derived from first principles of Newtonian mechanics, and that the wave motion is studied in the Boussinesq scaling which allows for both nonlinear and dispersive effects. The dynamics of the motion of the bottom have been developed following recent work in \cite{KhakimzyanovG.S.Shokina2010}. The Boussinesq model which has been utilised here allows for a dynamic bathymetry, and was derived in \cite{Wu1987}. The numerical method used in this paper is an extension of the method put forward in \cite{Barth1994, Barth2004}. 

The results presented in Section~\ref{Results} clearly show that dispersion may have a strong effect on the run-up and draw-down at the beaches, but it is not clear which of the two models (the shallow-water, or the Boussinesq model) paints a more realistic picture of the actual wave conditions. We have no way of quantifying the energy dissipation in the shallow-water solver other than comparing the total wave energy with that of the Boussinesq model. As the difference is rather large, 
we expect a significant amount of numerical dissipation in the shallow-water simulation. 
While the shallow-water model simulation might be closer to physical
reality where actual damping occurs because of molecular viscosity and fluid-structure interactions, 
it is likely that the Boussinesq model exhibits a closer resemblance to the Euler
equations which are taken as the basic governing equations in this work.

Of course, this difference could be more or less pronounced depending on the particular case under study. For example, the divergence between the shallow-water theory and the dispersive model is stronger at the right beach than at the left beach. The results also show that a finite domain exhibits different behavior than a half-open domain (such as used in \cite{Grilli2005}) with respect to the dependence of the wave run-up on the initial depth of the landslide. While the run-up is a strictly decreasing function of the initial depth in an open domain, a closed domain appears to exhibit resonant effects, which make the dependence more complex.

\section*{Acknowledgements}

D.~Dutykh would like to thank the University of Bergen for support and hospitality during the preparation of this manuscript. Support from the Agence Nationale de la Recherche under project ANR-08-BLAN-0301-01 (MathOc\'{e}an) and from ERC under project ERC-2011-AdG 290562-MULTIWAVE is also gratefully acknowledged.

H.~Kalisch acknowledges support of the Research Council of Norway through grant no. NFR 213474/F20.

\end{document}